\documentclass[a4paper,11pt]{article}
\usepackage[inline]{enumitem}
\usepackage[multidot]{grffile} 
\usepackage{dcolumn}
\usepackage{bm}
\usepackage{amsmath}
\usepackage{amsthm}
\usepackage{framed}
\usepackage{soul}
\usepackage{amsfonts}
\usepackage{bbold}
\usepackage{dsfont}
\usepackage{amssymb}
\usepackage{color}
\usepackage{latexsym}
\usepackage{stackrel}
\usepackage{slashed} 
\usepackage{empheq}
\usepackage{fancybox}
\usepackage{pstricks}
\usepackage{indentfirst}
\usepackage{mathrsfs}
\usepackage{tablefootnote}
\usepackage{longtable}
\usepackage{multirow}
\usepackage{epsfig,psfrag}
\usepackage{subfigure}
\usepackage{mathtools}
\usepackage{setspace} 
\usepackage[utf8]{inputenc} 
\usepackage[scientific-notation=true]{siunitx} 
\usepackage{verbatim}
\usepackage{comment}
\usepackage[many]{tcolorbox} 
\usepackage{JHEPpub}
\usepackage{JHEPorcid} 
\usepackage[T1]{fontenc} 
\graphicspath{fig/}
\title{QED nuclear medium effects at EicC}
\author[a,b]{Oleksandr Tomalak\orcidlink{0000-0002-4827-5842}}
\emailAdd{tomalak@itp.ac.cn}
\affiliation[a]{Institute of Theoretical Physics, Chinese Academy of Sciences, Beijing 100190, China}
\affiliation[b]{University of Chinese Academy of Sciences, School of Physical Sciences, Beijing 100049, China}
\abstract{We evaluate quantum electrodynamics (QED) nuclear medium effects under experimental conditions at the future Electron-ion collider in China (EicC). We consider neutral-current and charged-current elastic scattering, as well as neutral-current and charged-current deep inelastic scattering. We compute cross-section corrections at first order in the opacity expansion and estimate kinematic modifications due to multiple rescattering within the nucleus. We perform the first calculations of charged-current deep inelastic scattering and polarized scattering. We present results for the anticipated $^{197}_{79}\mathrm{Au},~^{208}_{82}\mathrm{Pb},$ and $^{238}_{92}\mathrm{U}$ nuclei and extrapolate these calculations to the lighter $^{40}_{20}\mathrm{Ca}$ isotope. We find that QED nuclear medium effects are non-negligible when extracting the process-independent non-perturbative structure of nucleons and nuclei at the EicC.}


\begin{document}
\maketitle
%

\section{Introduction}

The scattering of electrons, muons, and (anti)neutrinos from nucleon and nuclear targets provides the most efficient path toward understanding complex internal dynamics at nuclear length scales and below. These data provide access to one-dimensional and multidimensional structure~\cite{Hofstadter:1956qs,Ernst:1960zza,Hand:1963zz,Dombey:1969wk,Akhiezer:1973xbf,Drechsel:1989ab,Boffi:1993gs,Blomqvist:1998xn,Leemann:2001dg,Arrington:2006zm,Perdrisat:2006hj,Zhan:2011ji,Abrahamyan:2012gp,Accardi:2012qut,Qweak:2013zxf,MUSE:2013uhu,Mosel:2016cwa,Aschenauer:2017jsk,Adams:2018pwt,Anderle:2021wcy,A1:2010nsl,A1:2013fsc,Meyer:2016oeg,Xiong:2019umf,Borah:2020gte,Perdrisat:2006hj,JeffersonLabHallA:2022cit,JeffersonLabHallA:2022ljj} and serve as essential input for precise investigations of Standard Model interactions~\cite{Erler:2004in,SLACE158:2005uay,Kumar:2013yoa,Becker:2018ggl,MINOS:2011amj,T2K:2011qtm,Hyper-KamiokandeProto-:2015xww,T2K:2019bcf,NOvA:2019cyt,DUNE:2020ypp} and searches for new physics.

Achieving precision in scattering experiments with lepton probes and accurately extracting the structure-dependent information, encoded in matrix elements of quantum chromodynamics (QCD) operators, is feasible only after unfolding quantum electrodynamics (QED) radiative corrections. For this reason, dedicated studies of QED interactions across various energy scales~\cite{Yennie:1961ad,Mo:1968cg,Maximon:2000hm,Vanderhaeghen:2000ws,Liu:2020rvc,Afanasev:2023gev,Crowe:2026lky,DeRujula:1979grv,Day:2012gb,Tomalak:2021hec,Tomalak:2022xup} and the implementation of these results in event generators~\cite{Gramolin:2014pva,Banerjee:2020rww,Campbell:2022qmc,TenaVidal:2024eyt} represent a well-developed and active research field.

QED interactions with a medium at microscopic scales have been investigated for several decades. The suppression of bremsstrahlung due to quantum-mechanical interference from the surrounding medium is known as the Landau--Pomeranchuk--Migdal (LPM) effect~\cite{Landau:1953um,Landau:1953gr,Migdal:1956tc}. Modern quantum-field-theoretic formulations were developed in~\cite{Baier:1996vi,Zakharov:1996fv,Arnold:2018fjr}, while the extension to include the pair-production channel has been recently addressed in~\cite{Arnold:2025dqj,Arnold:2026lwd,Arnold:2026wjy}. The LPM effects have been verified experimentally in~\cite{Kasahara:1985ke,Anthony:1995fs,SLAC-E-146:1997hnd,Hansen:2004ti,CERNNA63:2013ahd}. Moreover, the Coulomb corrections of order $\mathcal{O} \left( Z \alpha \right)$, with the nuclear charge $Z$ and the electromagnetic coupling constant $\alpha$, have been investigated in traditional~\cite{Calva-Tellez:1978ufm,Engel:1997fy,Tjon:2006qe} and recent~\cite{Kuraev:2013sea,Tuchin:2013eya,Hill:2023bfh} works. The effects on the transverse momentum distribution of an electron resulting from the exchange of multiple Coulomb photons with a nuclear charge distribution were also discussed in Refs.~\cite{Bertulani:1987tz,Vidovic:1992ik,Sun:2020ygb}, while the associated bremsstrahlung and energy loss were addressed in Refs.~\cite{Olsen:2003mj,Lee:2004ina} and Ref.~\cite{Sandrock:2018ivj}, respectively.

The increasing precision of electron scattering off nuclear targets~\cite{A1:2010nsl,A1:2013fsc,Xiong:2019umf,Willeke:2021ymc,AbdulKhalek:2021gbh,Anderle:2021wcy} raises the question of QED interactions with the electric charges inside the nucleus~\cite{Tomalak:2022kjd}. Based on developments in QCD nuclear medium modifications~\cite{Gyulassy:2002yv,Gyulassy:2000fs,Gyulassy:2000er,Wiedemann:2000za,Idilbi:2008vm,
Accardi:2009qv,Ovanesyan:2011xy,Rothstein:2016bsq,Barata:2020rdn}, we formulated and evaluated QED nuclear medium-induced cross-section corrections to elastic electron-nucleon and (anti)neutrino-nucleon scattering inside large nuclei at first order in the opacity expansion~\cite{Tomalak:2022kjd}. We found non-negligible corrections to electron cross sections at small scattering angles. Subsequently, we resummed multiple reinteractions along charged-lepton trajectories~\cite{Tomalak:2023kwl} and evaluated the QED nuclear medium-induced radiation~\cite{Tomalak:2024lme}. In Ref.~\cite{Bhattacharya:2025pje}, we performed the first evaluations of corresponding effects inside the $^{208}_{82}\mathrm{Pb}$ nucleus in neutral-current deep inelastic scattering (DIS) at the energies of the forthcoming electron-ion colliders in the USA and China~\cite{Willeke:2021ymc,AbdulKhalek:2021gbh,Anderle:2021wcy,Xiao:2026tbs}.

In this paper, we extend our previous studies to the charged-current inclusive DIS, polarized neutral-current inclusive DIS, and charged-current elastic electron scattering. For the anticipated EicC facility~\cite{Anderle:2021wcy}, we perform calculations for $^{40}_{20}\mathrm{Ca},~^{197}_{79}\mathrm{Au},~^{208}_{82}\mathrm{Pb},$ and $^{238}_{92}\mathrm{U}$ nuclei for the corresponding energy of EicC and investigate the dependence on the nuclear size and parton distribution functions.

The paper is organized as follows. In section~\ref{sec:one}, we present the soft-collinear effective field theory formalism for evaluating effects of the QED nuclear medium on the scattering cross sections of charged particles. In subsections~\ref{subsec:one_NC_elastic}-\ref{subsec:one_CC_inclusive_DIS}, we numerically calculate and present the resulting $\mathcal{O} \left( \alpha^2 \right)$ cross-section corrections for neutral-current elastic, charged-current elastic, unpolarized and polarized neutral-current inclusive deep inelastic, and charged-current inclusive deep inelastic electron-nucleus scattering under EicC experimental conditions. In section~\ref{sec:multiple}, we account for QED nuclear medium-induced broadening of charged lepton tracks. We evaluate corresponding cross-section corrections and present the results in subsections~\ref{subsec:multiple_NC_elastic}-\ref{subsec:multiple_CC_inclusive_DIS} for the processes above. Our conclusions and outlook are collected in section~\ref{sec:conclusions_and_outlook}. We detail the charged-current elastic electron-proton scattering cross section in appendix~\ref{app:elastic_xsec}. In appendix~\ref{app:inclusive_xsec}, we provide expressions for neutral-current and charged-current inclusive deep inelastic scattering cross sections on a single nucleon.

\section{First order in the opacity expansion}
\label{sec:one}

In this section, we describe the formalism for evaluating QED nuclear medium effects on scattering cross sections at leading order in $\alpha^2$, the first order in the opacity expansion, and present the results for the corresponding modifications of neutral-current and charged-current scattering, unpolarized and polarized scattering, and elastic and deep inelastic electron scattering inside $^{40}_{20}\mathrm{Ca},~^{197}_{79}\mathrm{Au},~^{208}_{82}\mathrm{Pb},$ and $^{238}_{92}\mathrm{U}$ nuclei for the anticipated energy of the future EicC experiment~\cite{Anderle:2021wcy}.

Traveling through the nuclear medium, electrically charged particles interact electromagnetically with charges inside the nucleus, predominantly with protons. To quantify rescattering of high-energy electrons with energies above $\sim$10 MeV, it is convenient to adopt the collinear description. In the no-recoil approximation, QED interactions with the nuclear medium are described by the Coulomb potential $v$, mediated by the exchange of Glauber photons~\cite{Idilbi:2008vm,Ovanesyan:2011xy,Rothstein:2016bsq,Tomalak:2022kjd,Tomalak:2023kwl} having an enhanced momentum component $\vec{q}_\perp$ perpendicular to the electron three-momentum. The Coulomb potential in momentum space is expressed as
\begin{equation}
	v \left( \vec{q}_\perp \right) = \frac{4 \pi \alpha}{\vec{q}^2_\perp + \zeta^2}, \label{eq:Coulomb_potential}
\end{equation}
where the regularization at large distances is determined by the atomic screening scale $\zeta \approx \frac{n^2 m_e Z^{1/3}}{192} $~\cite{Jackson:1998nia}, with the nuclear charge $Z$, the principal quantum number $n$, and the electron mass $m_e$. To estimate uncertainties associated with the choice of the scale $\zeta$, we perform evaluations for the innermost and outermost electron orbits corresponding to $n=1$ and $n=n_\mathrm{max}$, respectively.\footnote{In this work, we use $n_\mathrm{max} \left( ^{40}_{20}\mathrm{Ca} \right) = 4$, $n_\mathrm{max} \left( ^{197}_{79}\mathrm{Au} \right) = 6$, $n_\mathrm{max} \left( ^{208}_{82}\mathrm{Pb} \right) = 6$, and $n_\mathrm{max} \left( ^{238}_{92}\mathrm{U} \right) = 7$.} The exchange of Glauber photons preserves the collinear description of the electron, resulting in relatively small electron deflections inside the nucleus~\cite{Tomalak:2023kwl}. However, the exchange of Glauber photons can substantially modify scattering cross sections.

At first order in the opacity expansion~\cite{Gyulassy:2000fs,Gyulassy:2000er,Wiedemann:2000za,Ovanesyan:2011xy}, we sum all contributions involving one and two exchanges of Glauber photons between the electron and the nuclear medium and obtain the unpolarized cross-section correction $\delta \sigma_e$~\cite{Tomalak:2022kjd},\footnote{For charged-current electron scattering, the first term in Eq.~(\ref{eq:QED_medium_electron}) is absent.}
\begin{eqnarray}
	\delta \sigma_e &=& \int \frac{\mathrm{d}^2 \vec{q}_{\perp}}{\left( 2 \pi \right)^2} |v \left( \vec{q}_\perp \right)|^2 \Bigg\{ \int \mathrm{d} z^\prime \rho \left( z^\prime \right) \left[ \sigma_e \left( \vec p \thinspace ^\prime - \vec{q}_\perp,\vec{p} \right) - \sigma_e \left( \vec p \thinspace ^\prime,\vec{p} \right) \right] \nonumber \\ 
	&& \hspace*{3cm}+ \int \mathrm{d} z \rho \left( z \right) \left[ \sigma_e \left( \vec p \thinspace ^\prime, \vec{p} + \vec{q}_\perp \right) - \sigma_e \left( \vec p \thinspace ^\prime,\vec{p} \right) \right] \Bigg\}, \label{eq:QED_medium_electron}
\end{eqnarray}
to the medium-free hard scattering cross section $\sigma_e \left( \vec p \thinspace ^\prime,\vec{p} \right)$, where $\vec{p}$ and $\vec p \thinspace ^\prime$ are incoming and outgoing electron momenta and $p$ and $p^\prime$ are the corresponding four-momenta. The integration in coordinate space is performed along the trajectories of the incoming ($z$) and outgoing ($z^\prime$) electrons, weighted by the nuclear charge density $\rho$, taken along the trajectory. The momentum-space integration is performed over the transverse momentum of the Glauber photon $\vec{q}_\perp$ in the soft interaction.\footnote{We specify the soft scattering as the process with the smaller squared momentum transfer.} In contrast to well-known Coulomb effects~\cite{Engel:1997fy,Tjon:2006qe,Bertulani:1987tz,Vidovic:1992ik,Olsen:2003mj,Lee:2004ina,Sandrock:2018ivj,Sun:2020ygb}, the correction $\delta \sigma_e$ starts at $\alpha^2$ and can become substantial due to the logarithmic enhancement from the ratio of the atomic scale to the experimental scale. Therefore, we can safely neglect the medium-induced contribution arising from the charge distribution of atomic electrons and instead absorb it into the uncertainty estimate. In this paper, we assume that scattering occurs on individual nucleons and neglect nucleon motion inside nuclei, correlations among initial-state nucleons, and final-state interactions. We also do not consider medium-induced radiation, which results in parametrically and numerically suppressed effects~\cite{Tomalak:2024lme}.

For numerical estimates, we use the Woods-Saxon spherically symmetric distribution for protons inside the nucleus,
\begin{equation}
	\rho_0 \left( r \right) = -\frac{1}{8 \pi a^3 \mathrm{Li}_3 \left( - e^{\frac{R_0}{a}} \right) } \frac{Z}{1 + e^\frac{r-R_0}{a}}, \label{eq:Woods_Saxon}
\end{equation}
with the nucleus size parameter $R_0$, which is taken according to the experimental data for nuclear charge radii in Ref.~\cite{nds_charge_radii}, and the fixed constant $a = 0.5~\mathrm{fm}$. For evaluations in this section, we average over all nucleons as potential scattering centers inside the nucleus and add both contributions from the scattering off the neutron and off the proton. For scattering off neutrons, we use the same expression normalized to the number of neutrons inside the nucleus.

At the high energy of EicC, the relative correction saturates and becomes independent of the beam energy. In this regime, calculations using the soft-collinear effective field theory are in perfect agreement with exact QED results~\cite{Tomalak:2022kjd}.

\subsection{Neutral-current elastic scattering} \label{subsec:one_NC_elastic}

\begin{figure}[tb!]
	\centering
	{\includegraphics[angle=0,scale=0.42]{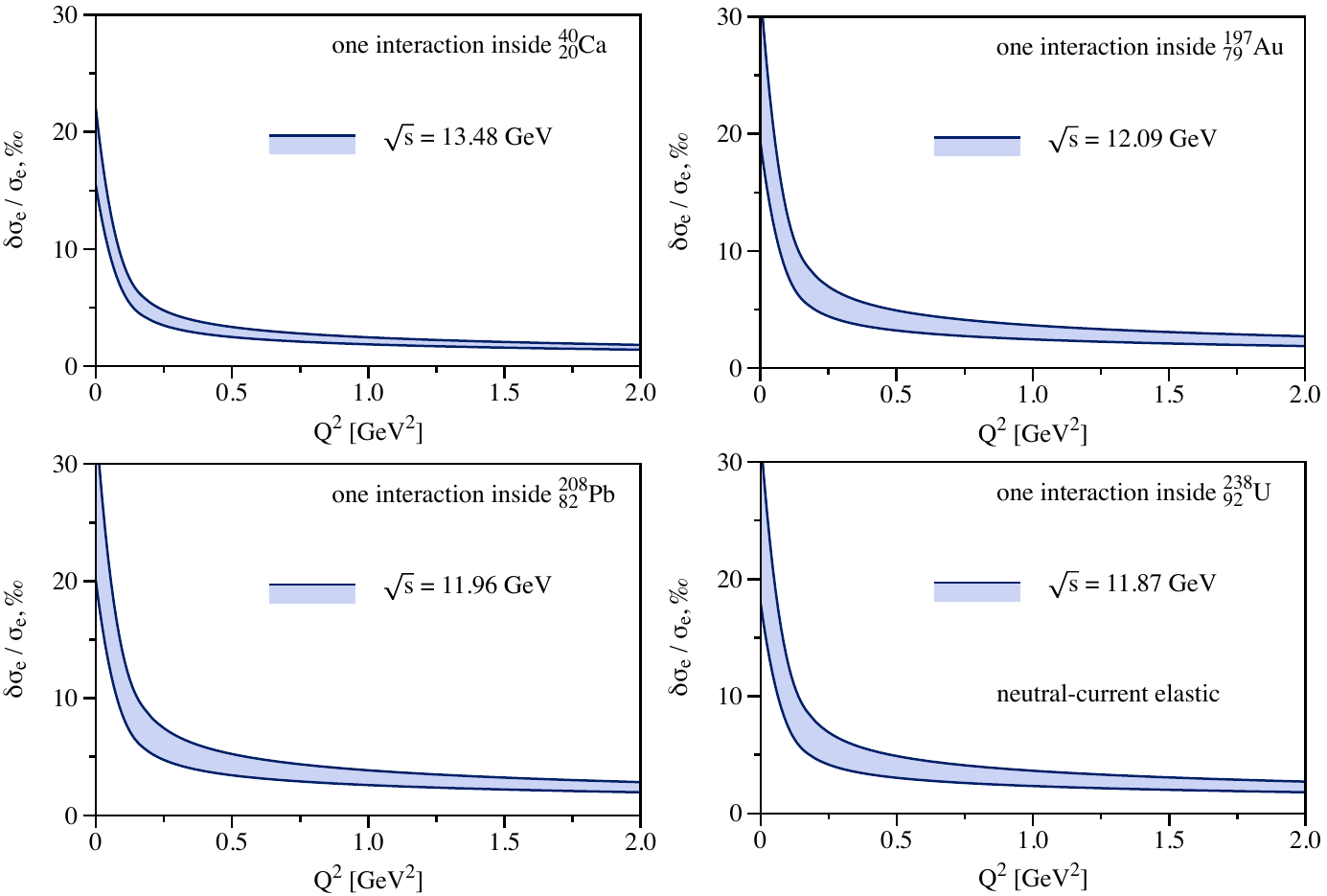}}
	\caption{Relative correction to neutral-current elastic electron-nucleus scattering cross section from QED nuclear medium effects inside $^{40}_{20}\mathrm{Ca},~^{197}_{79}\mathrm{Au},~^{208}_{82}\mathrm{Pb},$ and $^{238}_{92}\mathrm{U}$ nuclei at first order in the opacity expansion is shown as a function of the squared momentum transfer $Q^2$ for the electron beam energy of the future EicC. The upper and lower curves correspond to the choice of the atomic scale $\zeta =\frac{m_e Z^{\frac{1}{3}}}{192}$ and $\zeta =\frac{n_\mathrm{max}^2 m_e Z^{\frac{1}{3}}}{192}$, with the smallest and largest principal quantum numbers $1$ and $n_\mathrm{max}$, respectively.}
	\label{fig:one_NC_elastic}
	\centering
	{\includegraphics[angle=0,scale=0.42]{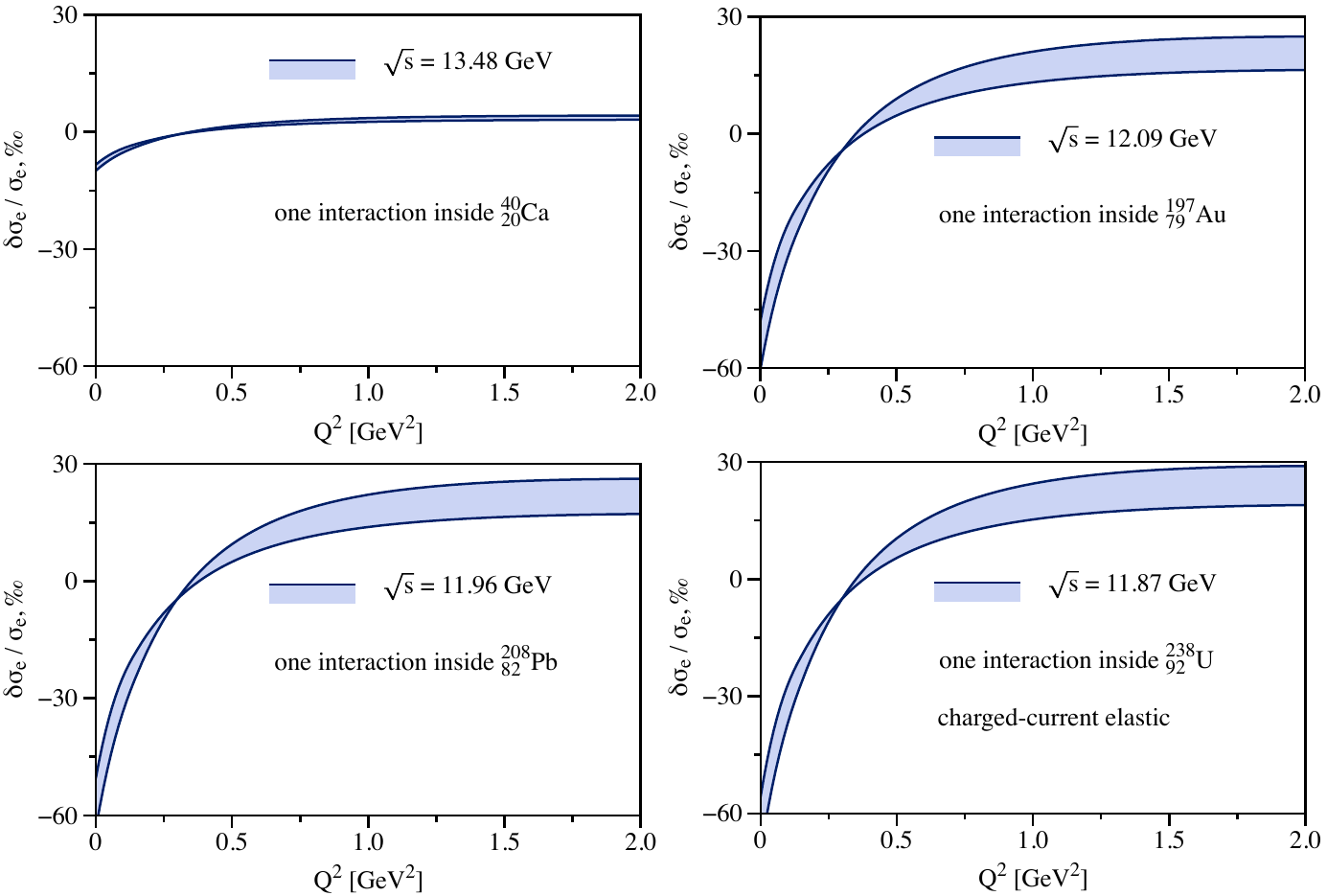}}
	\caption{Relative correction to the charged-current elastic electron-nucleus scattering cross section from QED nuclear medium effects inside $^{40}_{20}\mathrm{Ca},~^{197}_{79}\mathrm{Au},~^{208}_{82}\mathrm{Pb},$ and $^{238}_{92}\mathrm{U}$ nuclei at first order in the opacity expansion is shown as a function of the squared momentum transfer $Q^2$ for the electron beam energy of the future EicC. The upper and lower curves correspond to the choice of the atomic scale $\zeta =\frac{m_e Z^{\frac{1}{3}}}{192}$ and $\zeta =\frac{n_\mathrm{max}^2 m_e Z^{\frac{1}{3}}}{192}$, with the smallest and largest principal quantum numbers $1$ and $n_\mathrm{max}$, respectively.}
	\label{fig:one_CC_elastic}
\end{figure}

We evaluate QED nuclear medium effects in neutral-current elastic electron scattering by integrating the cross-section expressions from the appendix of Ref.~\cite{Tomalak:2022kjd} using Eq.~(\ref{eq:QED_medium_electron}), with the nucleon electromagnetic form factors from Ref.~\cite{Borah:2020gte} as an input. We present the results for cross-section corrections from QED nuclear medium effects inside $^{40}_{20}\mathrm{Ca},~^{197}_{79}\mathrm{Au},$ $^{208}_{82}\mathrm{Pb},$ and $^{238}_{92}\mathrm{U}$ nuclei at first order in the opacity expansion as a function of the squared momentum transfer $Q^2 = - \left( p - p^\prime \right)^2$ in figure~\ref{fig:one_NC_elastic} for the anticipated EicC energy in the center-of-mass reference frame $s = \left( p + k \right)^2$. The resulting correction to neutral-current elastic electron scattering is positive definite. As in all previous evaluations~\cite{Tomalak:2022kjd,Bhattacharya:2025pje}, the correction is larger at small squared momentum transfers and has a comparable magnitude for the heavy nuclei of interest. Extrapolations to the lighter $^{40}_{20}\mathrm{Ca}$ nucleus indicate a smaller correction, with a suppression factor of less than $1.5$. The relative correction exceeds $1\%$ in magnitude for squared momentum transfers $Q^2 \lesssim 0.2~\mathrm{GeV}^2$.

\subsection{Charged-current elastic scattering} \label{subsec:one_CC_elastic}

Charged-current electron scattering might provide important input for describing scattering of (anti)neutrinos from atomic nuclei and for constraining the matrix elements of the axial-vector currents. According to sensitivity studies in~\cite{JLab-PR12-25-009,Klest:2025bfl,Yang:2026vuf,Davoudiasl:2025ifk,Fatima:2026mac,Fatima:2026hyc}, such measurements are potentially feasible with current and future electron and positron beams.

We evaluate QED nuclear medium effects in charged-current elastic electron scattering by integrating the cross-section expressions from appendix~\ref{app:elastic_xsec} with the incoming electron trajectory in the second line of Eq.~(\ref{eq:QED_medium_electron}), using the nucleon vector form factors from Ref.~\cite{Borah:2020gte} and the axial-vector form factor from Ref.~\cite{Meyer:2016oeg} as input. We present the results of numerical integration inside $^{40}_{20}\mathrm{Ca},~^{197}_{79}\mathrm{Au},~^{208}_{82}\mathrm{Pb},$ and $^{238}_{92}\mathrm{U}$ nuclei in figure~\ref{fig:one_CC_elastic}. In contrast to neutral-current elastic electron scattering, the QED nuclear medium effects at first order in the opacity expansion exhibit a sign change in charged-current elastic electron scattering, thereby distorting the medium-free cross section. The squared momentum transfer dependence of the relative correction is monotonic. This reflects the dependence observed in charged-current elastic antineutrino scattering~\cite{Tomalak:2022kjd}, where the QED nuclear medium-induced contribution is negative at small squared momentum transfer and subsequently changes sign at larger $Q^2$. The interaction with the medium in electron-induced reactions, driven by soft-photon exchanges from the incoming rather than the outgoing electron, results in larger effects than in neutrino-induced scattering. The relative QED nuclear medium effects for heavy nuclei are at the few-percent level. For light nuclei, we find smaller corrections, reaching the percent level only at small squared momentum transfer.

\subsection{Neutral-current inclusive deep inelastic scattering} \label{subsec:one_NC_inclusive_DIS}

To describe inclusive deep inelastic scattering (DIS), we use the Bjorken scaling variable $x = \frac{Q^2}{2 k \cdot \left( p - p^\prime \right)}$, where $k$ is the incoming nucleon four-momentum, as the second independent Lorentz-invariant variable. We evaluate QED nuclear medium corrections to the single-nucleon neutral-current inclusive DIS cross sections by integrating the tree-level unpolarized cross-section expressions and the differences between cross sections with electron polarization along and opposite to the beam direction using Eq.~(\ref{eq:QED_medium_electron}), with the baseline cross sections taken from appendix~\ref{app:inclusive_xsec}. We use the Mathematica package ManeParse for parton distribution functions (PDFs)~\cite{Clark:2016jgm} for numerical inputs, with the bound-proton fit for the corresponding nucleus~\cite{Kovarik:2015cma,Kusina:2020lyz} and obtain the neutron PDFs under the assumption of $\mathrm{SU} \left( 2 \right)$ isospin symmetry.\footnote{In the absence of fits for scattering off $^{238}_{92}\mathrm{U}$, we perform the $^{238}_{92}\mathrm{U}$ calculation using the parton distribution functions of the heaviest available nucleus, $^{208}_{82}\mathrm{Pb}$. Comparing the calculations obtained with PDFs for different nuclei, we find only a mild dependence on the input PDFs outside the regions of small $x$ and small $Q^2$.}

\begin{figure}[htb!]
	\centering
	{\includegraphics[angle=0,scale=0.39]{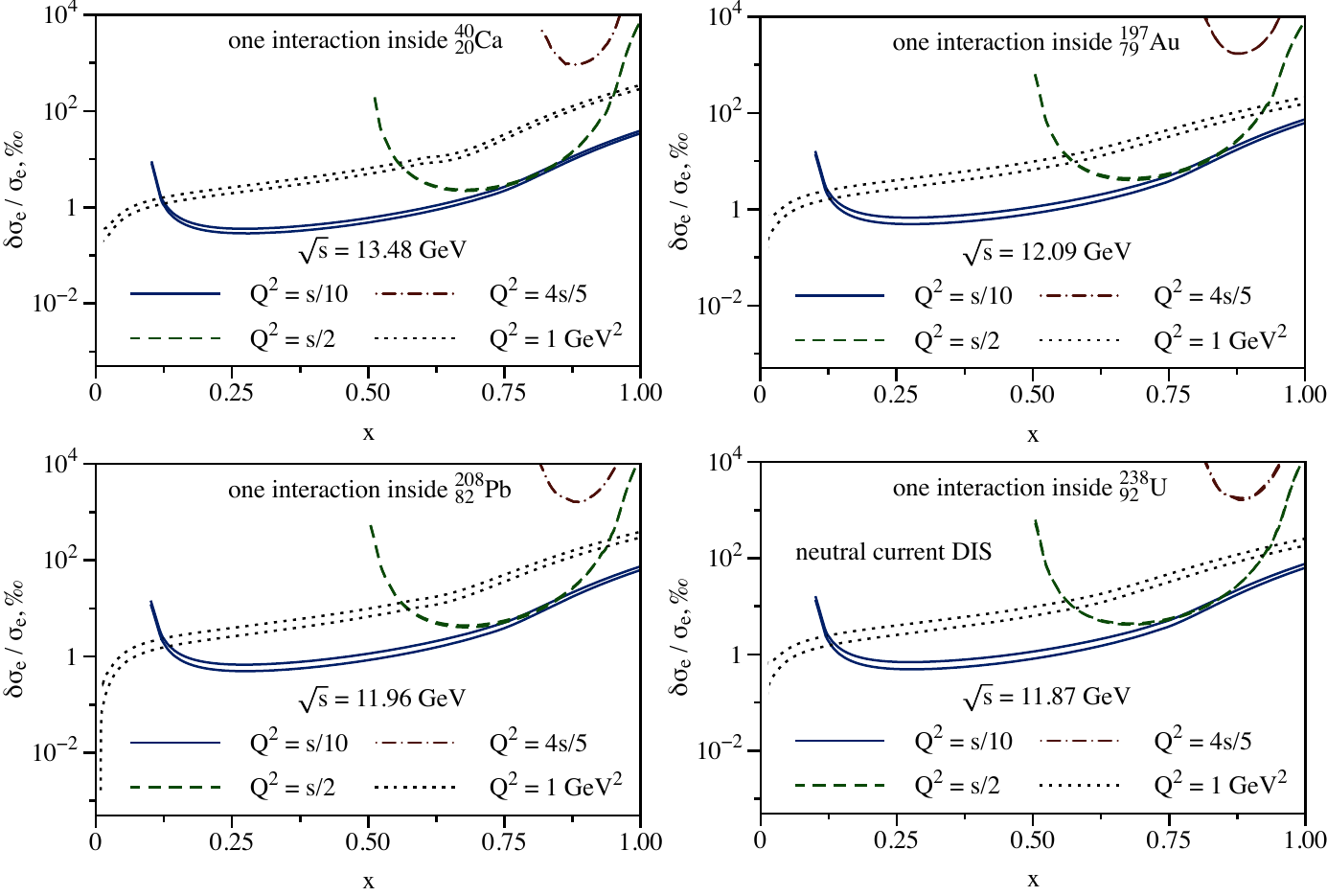}}
	\caption{Relative correction to the unpolarized neutral-current inclusive deep inelastic electron-nucleus scattering cross section from QED nuclear medium effects inside $^{40}_{20}\mathrm{Ca},~^{197}_{79}\mathrm{Au},~^{208}_{82}\mathrm{Pb},$ and $^{238}_{92}\mathrm{U}$ nuclei at first order in the opacity expansion is shown as a function of the Bjorken variable $x$ for the electron beam energy of the future EicC and fixed values of the squared momentum transfer $Q^2 = 1~\mathrm{GeV}^2,~s/10,~s/2,$ and $4s/5$. The upper and lower curves correspond to the choice of the atomic scale $\zeta =\frac{m_e Z^{\frac{1}{3}}}{192}$ and $\zeta =\frac{n_\mathrm{max}^2 m_e Z^{\frac{1}{3}}}{192}$, with the smallest and largest principal quantum numbers $1$ and $n_\mathrm{max}$, respectively. Corrections induced by incoming and outgoing electrons are included.}
	\label{fig:one_NC_inclusive_DIS_x_unpolarized}
	\centering
	{\includegraphics[angle=0,scale=0.39]{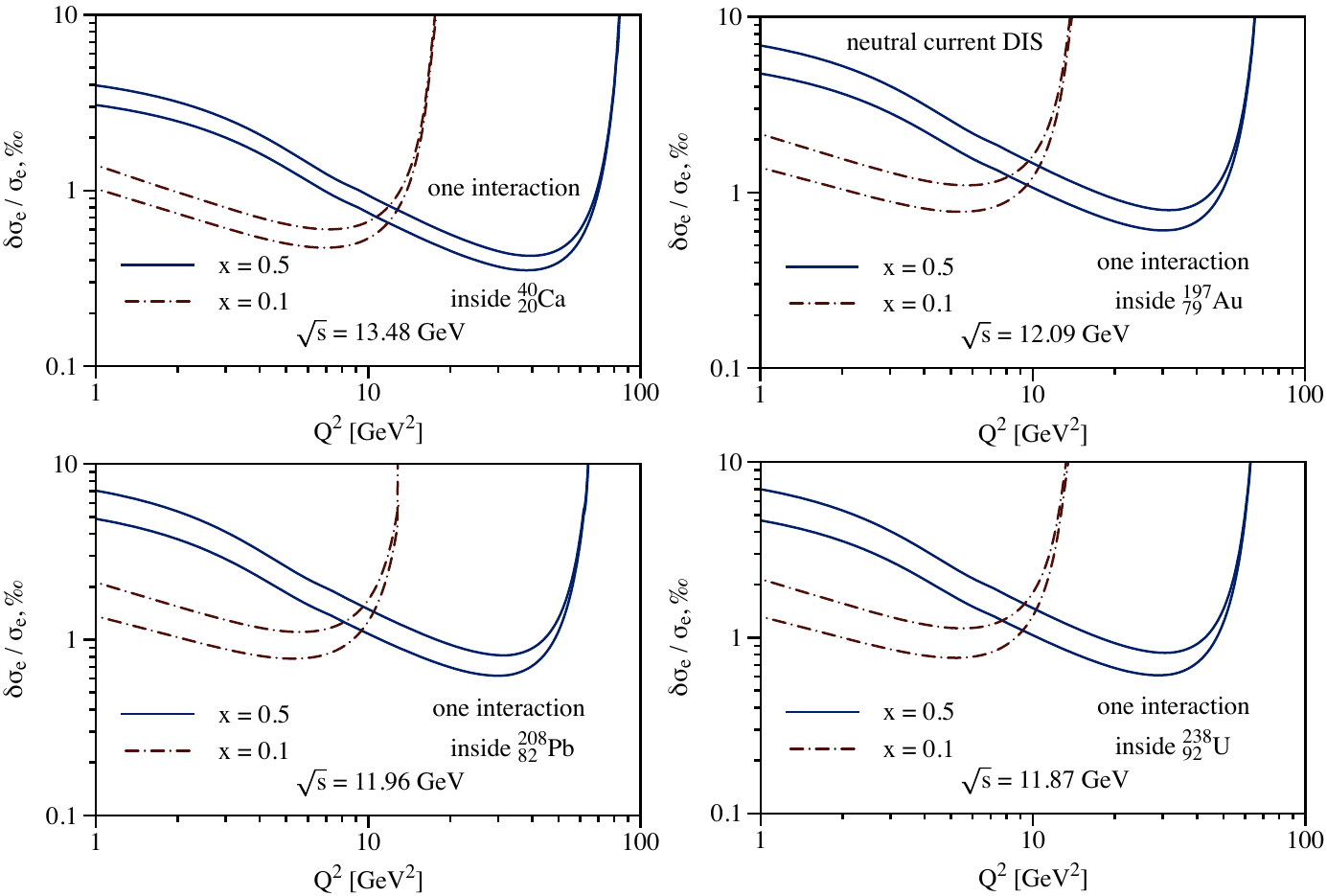}}
	\caption{Relative correction to the unpolarized neutral-current inclusive deep inelastic electron-nucleus scattering cross section from QED nuclear medium effects inside $^{40}_{20}\mathrm{Ca},~^{197}_{79}\mathrm{Au},~^{208}_{82}\mathrm{Pb},$ and $^{238}_{92}\mathrm{U}$ nuclei at first order in the opacity expansion is shown as a function of the squared momentum transfer $Q^2$ for the electron beam energy of the future EicC and fixed values of the Bjorken variable $x = 0.1,$ and $0.5$. The upper and lower curves correspond to the choice of the atomic scale $\zeta =\frac{m_e Z^{\frac{1}{3}}}{192}$ and $\zeta =\frac{n_\mathrm{max}^2 m_e Z^{\frac{1}{3}}}{192}$, with the smallest and largest principal quantum numbers $1$ and $n_\mathrm{max}$, respectively. Corrections induced by incoming and outgoing electrons are included.}
	\label{fig:one_NC_inclusive_DIS_Q2_unpolarized}
\end{figure}

We present the results of numerical integration for the unpolarized cross-section correction inside $^{40}_{20}\mathrm{Ca},~^{197}_{79}\mathrm{Au},~^{208}_{82}\mathrm{Pb},$ and $^{238}_{92}\mathrm{U}$ nuclei at fixed values of the squared momentum transfer $Q^2 = 1~\mathrm{GeV}^2,~s/10,~s/2,$ and $4s/5$ as a function of the Bjorken variable $x$ in figure~\ref{fig:one_NC_inclusive_DIS_x_unpolarized}. QED nuclear medium effects in neutral-current inclusive DIS increase with nuclear size and are similar for all three heavy nuclei $^{197}_{79}\mathrm{Au},~^{208}_{82}\mathrm{Pb},$ and $^{238}_{92}\mathrm{U}$ at the EicC. The relative corrections can reach the percent level only at large squared momentum transfer, large $x$, or near the kinematic boundaries of the Bjorken variable $x$.

In figure~\ref{fig:one_NC_inclusive_DIS_Q2_unpolarized}, we present relative QED nuclear medium effects in unpolarized neutral-current inclusive deep inelastic electron-nucleus scattering as a function of the squared momentum transfer $Q^2$ for a fixed value of the Bjorken variable $x$. The relative cross-section correction increases at the lowest and largest momentum transfers $Q^2$ and is larger at large $x$. The correction increases with nuclear size and is found to be comparable for all three heavy nuclei $^{197}_{79}\mathrm{Au},~^{208}_{82}\mathrm{Pb},$ and $^{238}_{92}\mathrm{U}$ of interest.
\begin{figure}[htb!]
	\centering
	{\includegraphics[angle=0,scale=0.39]{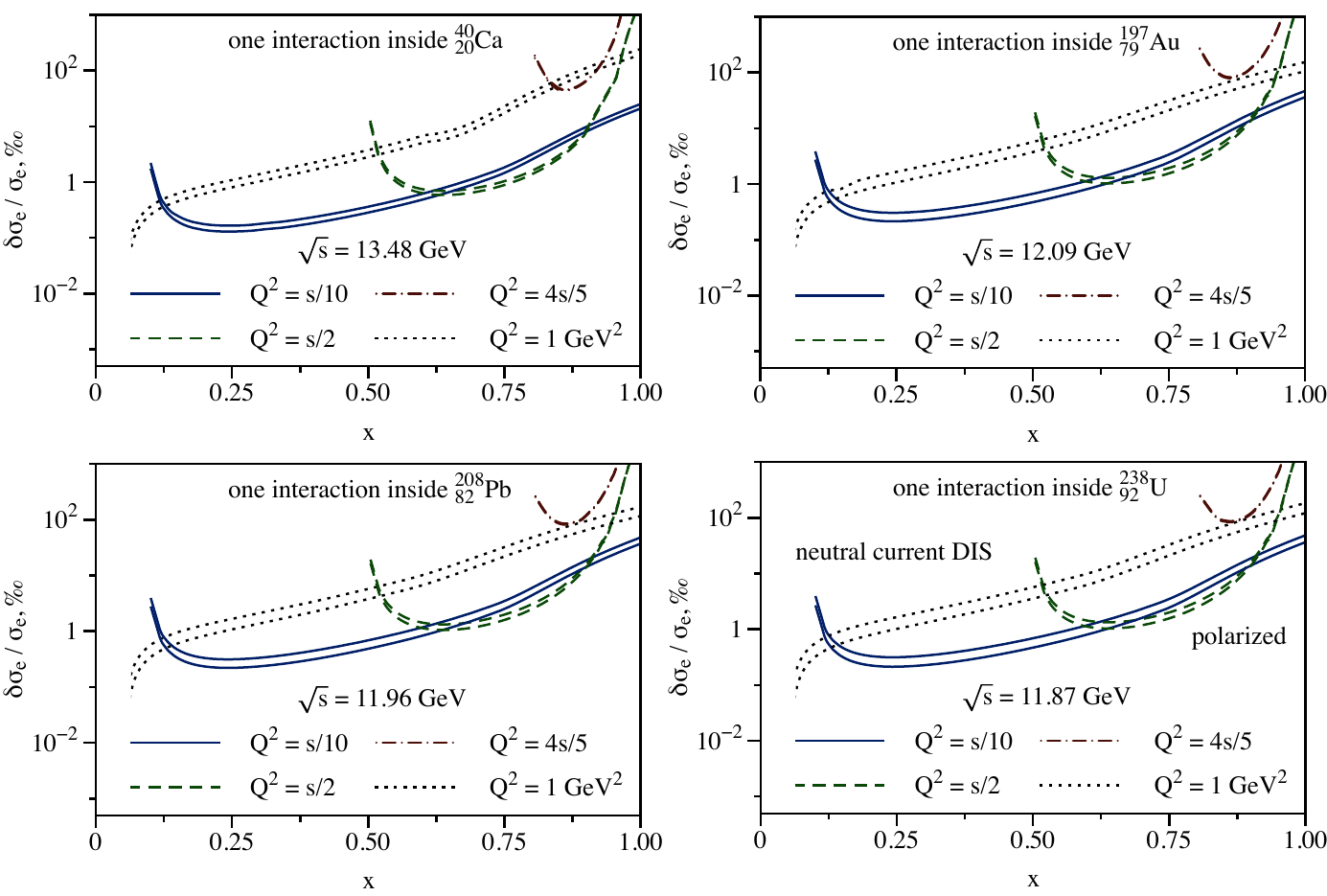}}
	\caption{Relative correction to the polarized neutral-current inclusive deep inelastic electron-nucleus scattering cross section from QED nuclear medium effects inside $^{40}_{20}\mathrm{Ca},~^{197}_{79}\mathrm{Au},~^{208}_{82}\mathrm{Pb},$ and $^{238}_{92}\mathrm{U}$ nuclei at first order in the opacity expansion is shown as a function of the Bjorken variable $x$ for the electron beam energy of the future EicC and fixed values of the squared momentum transfer $Q^2 = 1~\mathrm{GeV}^2,~s/10,~s/2,$ and $4s/5$. The cross section represents the single-spin asymmetry of linearly polarized electrons. The upper and lower curves correspond to the choice of the atomic scale $\zeta =\frac{m_e Z^{\frac{1}{3}}}{192}$ and $\zeta =\frac{n_\mathrm{max}^2 m_e Z^{\frac{1}{3}}}{192}$, with the smallest and largest principal quantum numbers $1$ and $n_\mathrm{max}$, respectively. Corrections induced by incoming and outgoing electrons are included.}
	\label{fig:one_NC_inclusive_DIS_x_polarized}
	\centering
	{\includegraphics[angle=0,scale=0.39]{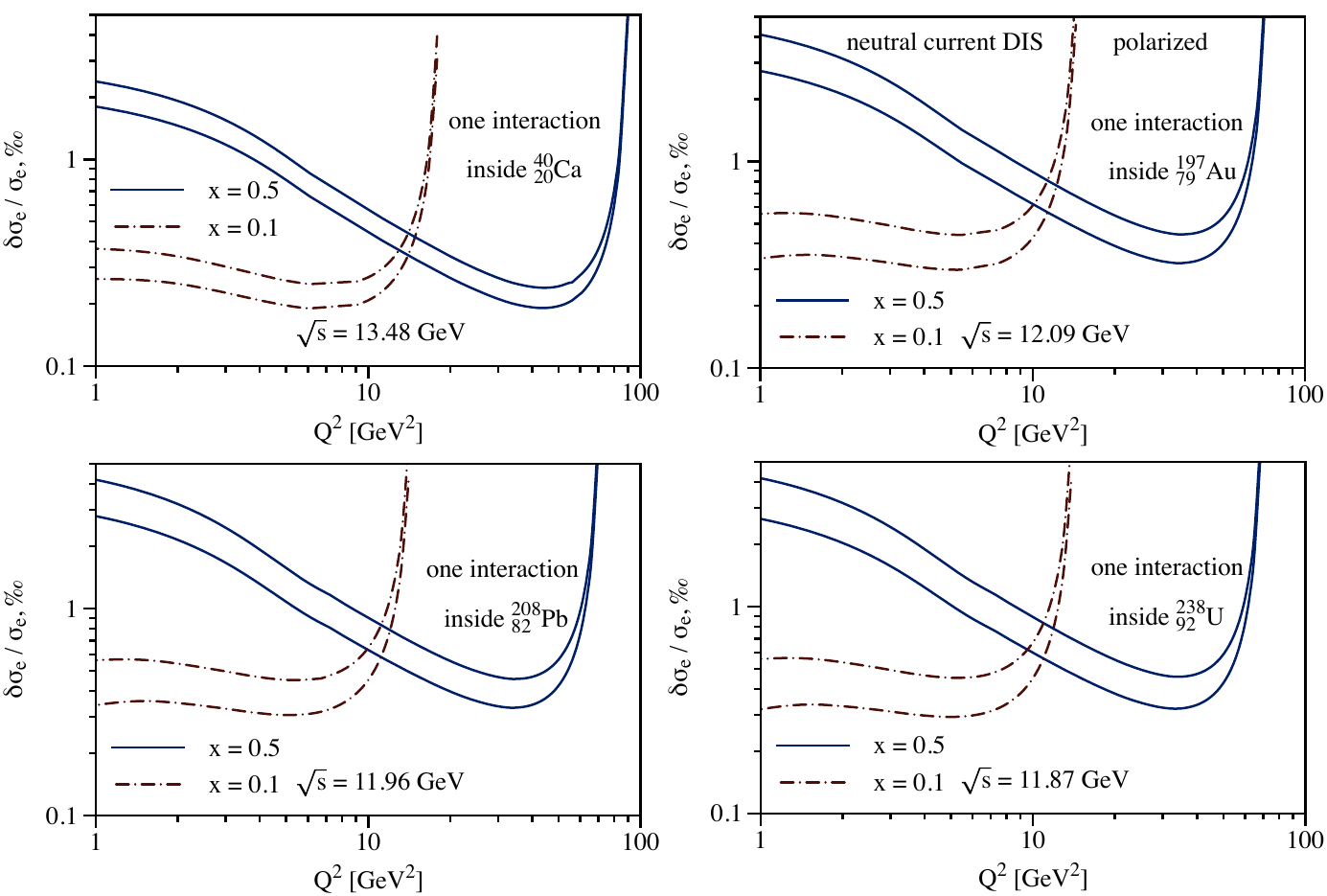}}
	\caption{Relative correction to the polarized neutral-current inclusive deep inelastic electron-nucleus scattering cross section from QED nuclear medium effects inside $^{40}_{20}\mathrm{Ca},~^{197}_{79}\mathrm{Au},~^{208}_{82}\mathrm{Pb},$ and $^{238}_{92}\mathrm{U}$ nuclei at first order in the opacity expansion is shown as a function of the squared momentum transfer $Q^2$ for the electron beam energy of the future EicC and fixed values of the Bjorken variable $x = 0.1,$ and $0.5$. The cross section represents the single-spin asymmetry of linearly polarized electrons. The upper and lower curves correspond to the choice of the atomic scale $\zeta =\frac{m_e Z^{\frac{1}{3}}}{192}$ and $\zeta =\frac{n_\mathrm{max}^2 m_e Z^{\frac{1}{3}}}{192}$, with the smallest and largest principal quantum numbers $1$ and $n_\mathrm{max}$, respectively. Corrections induced by incoming and outgoing electrons are included.}
	\label{fig:one_NC_inclusive_DIS_Q2_polarized}
\end{figure}
\begin{figure}[htb!]
	\centering
	{\includegraphics[angle=0,scale=0.39]{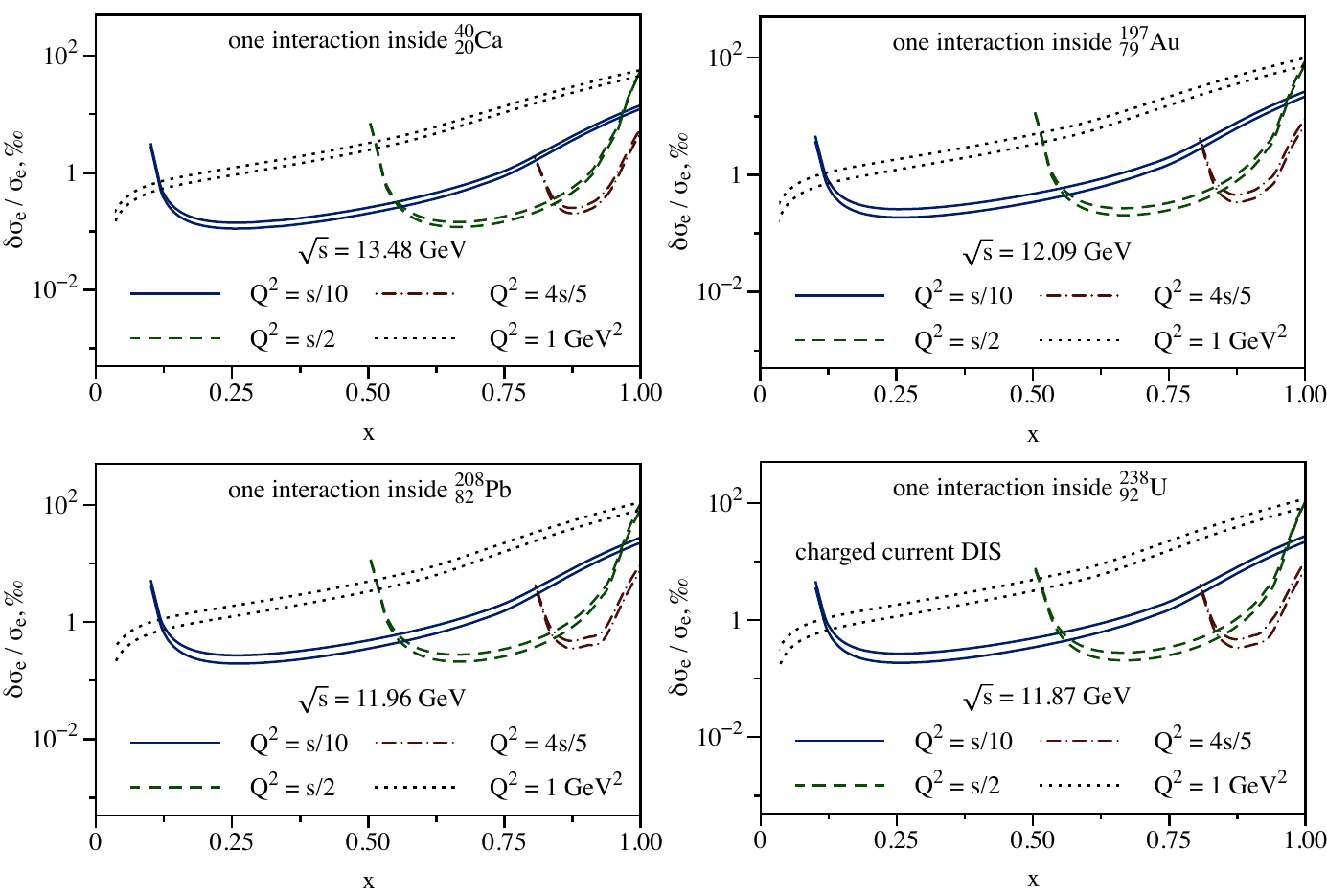}}
	\caption{Relative correction to the unpolarized charged-current inclusive deep inelastic electron-nucleus scattering cross section from QED nuclear medium effects inside $^{40}_{20}\mathrm{Ca},~^{197}_{79}\mathrm{Au},~^{208}_{82}\mathrm{Pb},$ and $^{238}_{92}\mathrm{U}$ nuclei at first order in the opacity expansion is shown as a function of the Bjorken variable $x$ for the electron beam energy of the future EicC and fixed values of the squared momentum transfer $Q^2 = 1~\mathrm{GeV}^2,~s/10,~s/2,$ and $4s/5$. The upper and lower curves correspond to the choice of the atomic scale $\zeta =\frac{m_e Z^{\frac{1}{3}}}{192}$ and $\zeta =\frac{n_\mathrm{max}^2 m_e Z^{\frac{1}{3}}}{192}$, with the smallest and largest principal quantum numbers $1$ and $n_\mathrm{max}$, respectively.}
	\label{fig:one_CC_inclusive_DIS_x_unpolarized}
	\centering
	{\includegraphics[angle=0,scale=0.39]{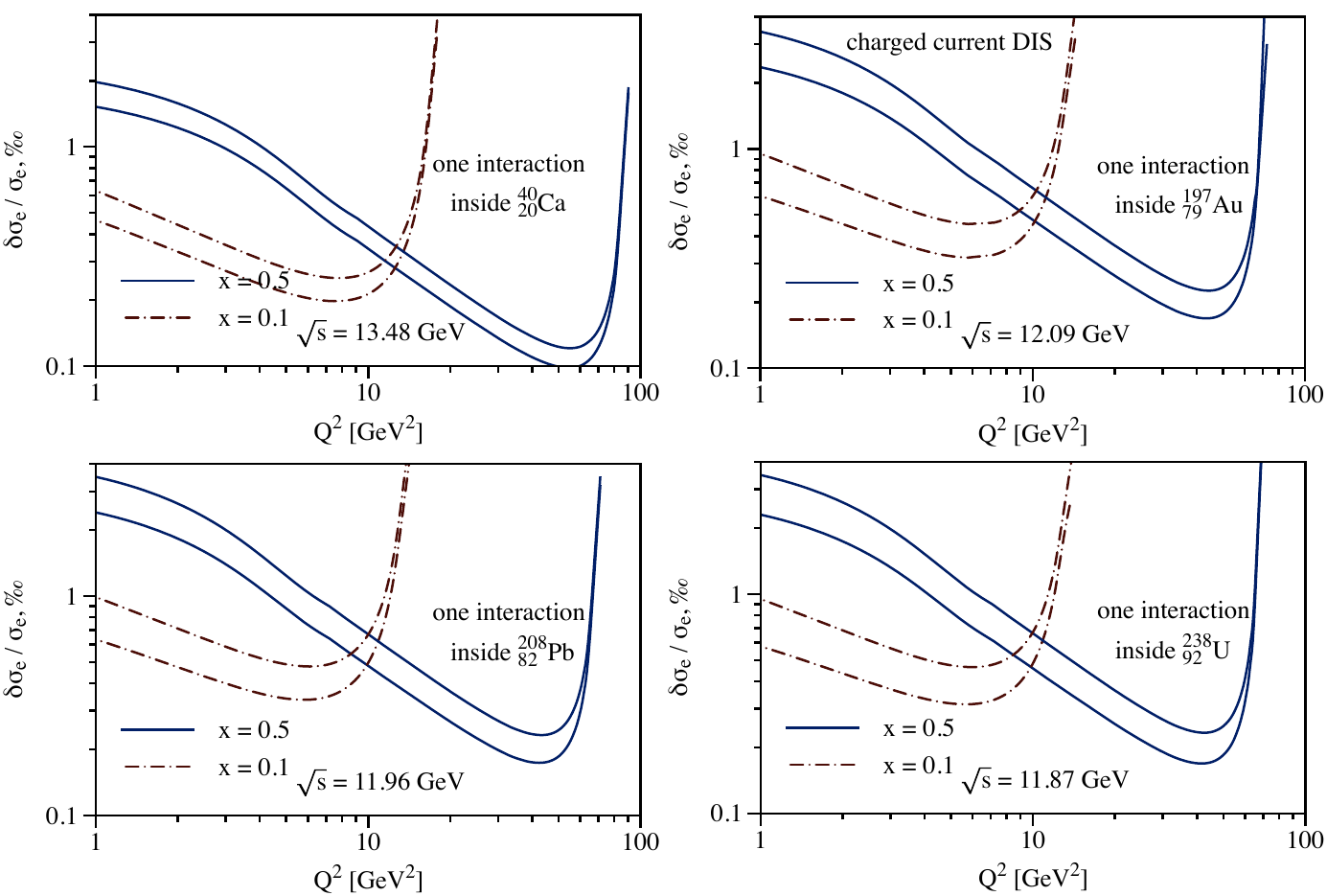}}
	\caption{Relative correction to the unpolarized charged-current inclusive deep inelastic electron-nucleus scattering cross section from QED nuclear medium effects inside $^{40}_{20}\mathrm{Ca},~^{197}_{79}\mathrm{Au},~^{208}_{82}\mathrm{Pb},$ and $^{238}_{92}\mathrm{U}$ nuclei at first order in the opacity expansion is shown as a function of the squared momentum transfer $Q^2$ for the electron beam energy of the future EicC and fixed values of the Bjorken variable $x = 0.01,~0.1,$ and $0.5$. The upper and lower curves correspond to the choice of the atomic scale $\zeta =\frac{m_e Z^{\frac{1}{3}}}{192}$ and $\zeta =\frac{n_\mathrm{max}^2 m_e Z^{\frac{1}{3}}}{192}$, with the smallest and largest principal quantum numbers $1$ and $n_\mathrm{max}$, respectively.}
	\label{fig:one_CC_inclusive_DIS_Q2_unpolarized}
\end{figure}

The analogous relative corrections for the single-spin asymmetry cross section of linearly polarized electrons are shown in figures~\ref{fig:one_NC_inclusive_DIS_x_polarized} and~\ref{fig:one_NC_inclusive_DIS_Q2_polarized} at fixed $Q^2$ and fixed $x$, respectively. The dependence of QED nuclear medium effects on nuclear size and kinematics in polarized scattering is qualitatively the same as in unpolarized scattering. The magnitude of the correction, however, is smaller for the spin-dependent part of inclusive deep inelastic scattering. Such an observable in neutral-current elastic electron-nucleus scattering, which involves only electromagnetic interactions, is suppressed by the electron mass compared to the unpolarized cross section in subsection~\ref{subsec:one_NC_elastic} and is therefore not of immediate interest.

\subsection{Charged-current inclusive deep inelastic scattering} \label{subsec:one_CC_inclusive_DIS}

To evaluate QED nuclear medium effects in charged-current inclusive DIS, we use the nucleon structure functions from appendix~\ref{app:inclusive_xsec} for interactions with the $W$ boson instead of those for the $\gamma$ and $Z$ bosons and employ the same inputs for PDFs~\cite{Clark:2016jgm,Kovarik:2015cma,Kusina:2020lyz} as in subsection~\ref{subsec:one_NC_inclusive_DIS}. We present our results at fixed $Q^2$ and fixed $x$ in figures~\ref{fig:one_CC_inclusive_DIS_x_unpolarized} and~\ref{fig:one_CC_inclusive_DIS_Q2_unpolarized}, respectively. The relative effects of the QED nuclear medium on charged-current cross sections with longitudinally polarized electrons are the same as in unpolarized scattering, while the cross section and QED nuclear medium effects vanish exactly for electrons with opposite polarization as a consequence of the chiral nature of charged-current electron interactions in ultrarelativistic scattering, when the electron mass can be safely neglected. The dependence of QED nuclear medium effects on nuclear size and kinematics in charged-current inclusive DIS is qualitatively the same as in unpolarized and polarized neutral-current inclusive DIS. The magnitude of corrections in charged-current scattering is smaller than in unpolarized neutral-current inclusive DIS and is comparable to that in polarized neutral-current DIS only at small values of $x$. However, QED nuclear medium effects in charged-current inclusive DIS at large $x$ are smaller than those in both polarized and unpolarized neutral-current inclusive DIS. In contrast to neutral-current inclusive DIS, the correction in charged-current inclusive DIS does not increase in magnitude with $x$ but maintains similar values, as is clearly visible in figure~\ref{fig:one_CC_inclusive_DIS_Q2_unpolarized}.

\section{Resummation of multiple rescattering}
\label{sec:multiple}

In this section, we evaluate effects from multiple QED rescattering within the nuclear medium.

Contributions to the unpolarized cross section from three or more Glauber-photon exchanges are suppressed by higher powers of $\alpha$. However, the electron trajectory can be significantly altered after a large number of soft interactions due to the transfer of perpendicular momentum from Glauber-photon exchanges~\cite{Ovanesyan:2011xy,Tomalak:2023kwl}. Multiple rescattering, predominantly in the forward direction, broadens the transverse momentum distribution of ultrarelativistic electrons ($p^\prime_\perp$) relative to the initial electron momentum with the following spectrum,\footnote{The transverse momentum distribution in Eq.~(\ref{eq:distribution_pT}) is normalized to unity.}
\begin{align} \label{eq:distribution_pT}
	\frac{\mathrm{d} N}{\mathrm{d} p^\prime_\perp} = \int \limits^{\infty}_{0} \mathrm{d} b \left( b p^\prime_\perp \right) J_0 \left( 0, b p^\prime_\perp \right) e^{\chi \left[ \left( \zeta b \right) K_1 \left( \zeta b \right) -1 \right]},
\end{align}
where the integration is performed over the radial coordinate $b$ in the transverse coordinate space, $K_1$ is the modified Bessel function of the second kind, $J_0$ is the Bessel function of the first kind, and $\chi$ denotes the mean number of QED interactions inside the nucleus with the r.m.s. radius $R_\mathrm{rms}$,
\begin{align} \label{eq:number_of_scatterings}
	\chi \sim \frac{Z^{1/3}}{\left(m_e R_\mathrm{rms} \right)^2}.
\end{align}
This number for scattering inside nuclei is rather large. For ultrarelativistic electrons, $\chi \sim 10^{5}-10^{7}$~\cite{Tomalak:2023kwl}, which motivates the resummation of multiple soft interactions. The resummation results in sizable transverse nucleon momenta of the order of $10$-$30~\mathrm{MeV}$ and can significantly distort the kinematics of the hard scattering process.

The broadening of the electron trajectory can significantly affect the reconstruction of kinematic invariants in the hard scattering process inside the nucleus. Using the elastic relation between the invariants and the recoil lepton energy $E^\prime_e$ and scattering angle, we determine the expected cross section as $\sigma^\mathrm{exp}$. On the other hand, the cross section in the experiment, the ``true" cross section $\sigma^\mathrm{broad}$, corresponds to averaging, using Eq.~(\ref{eq:distribution_pT}), the cross sections for all possible scattering angles in the hard interaction process over the transverse momentum distributions of initial- and final-state electrons. The effect of multiple rescattering on the unpolarized cross section can be quantified by the ratio $\sigma^\mathrm{broad}/\sigma^\mathrm{exp}$, which we compute numerically for the electron scattering cross section off a single proton inside the nucleus for the electron beam energy of the future EicC. We present our results in the following sections according to the order of the results in section~\ref{sec:one}.

QED nuclear medium rescattering decreases cross sections in all reactions in this section, i.e., neutral-current and charged-current elastic unpolarized electron-proton scattering, unpolarized and polarized neutral-current inclusive deep inelastic electron-proton scattering, and charged-current inclusive deep inelastic electron-proton scattering. The largest corrections are observed at small scattering angles. Unlike QED nuclear medium corrections at first order in the opacity expansion in section~\ref{sec:one}, which are sizable only in forward scattering, the relative effects are at the few-percent level over a wide range of kinematics.

\subsection{Neutral-current elastic scattering} \label{subsec:multiple_NC_elastic}

\begin{figure}[htb!]
	\centering
	{\includegraphics[angle=0,scale=0.42]{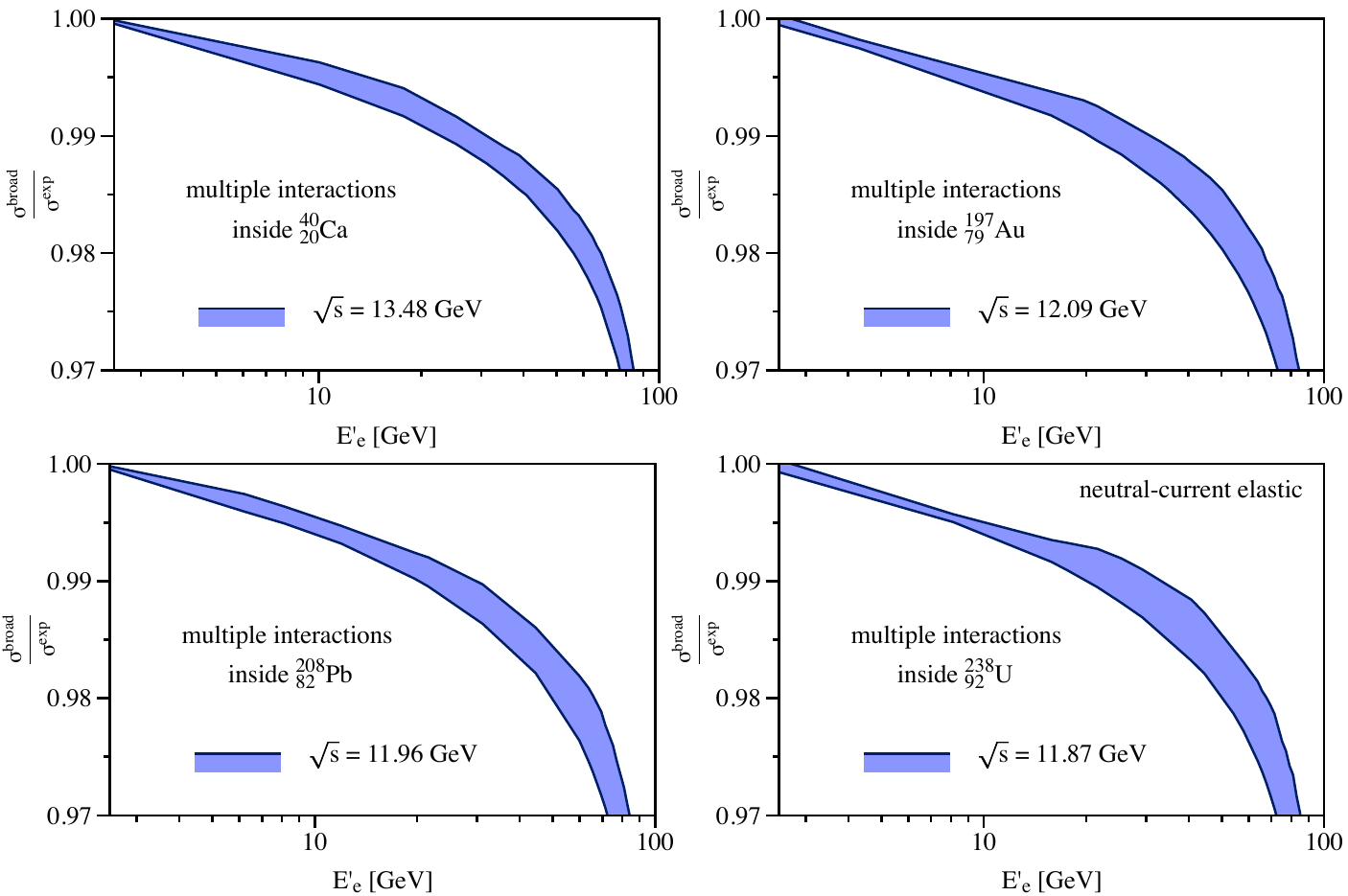}}
	\caption{Ratio of the neutral-current elastic electron-proton scattering cross section, after accounting for QED nuclear medium broadening of incoming and outgoing electrons inside $^{40}_{20}\mathrm{Ca},~^{197}_{79}\mathrm{Au},~^{208}_{82}\mathrm{Pb},$ and $^{238}_{92}\mathrm{U}$ nuclei, to the cross section evaluated from the observed lepton kinematics is shown as a function of the recoil electron energy $E_e^\prime$ in the rest frame of the initial proton for the electron beam energy of the future EicC. The lower and upper curves correspond to the choice of the atomic scale $\zeta =\frac{m_e Z^{\frac{1}{3}}}{192}$ and $\zeta =\frac{n_\mathrm{max}^2 m_e Z^{\frac{1}{3}}}{192}$, with the smallest and largest principal quantum numbers $1$ and $n_\mathrm{max}$, respectively.}
	\label{fig:multiple_NC_elastic}
\end{figure}
We average the unpolarized neutral-current elastic electron-proton scattering cross sections over the transverse momentum distributions of the initial and final electrons for all possible kinematic configurations of the hard interaction process inside $^{40}_{20}\mathrm{Ca},~^{197}_{79}\mathrm{Au},~^{208}_{82}\mathrm{Pb},$ and $^{238}_{92}\mathrm{U}$ nuclei and present the ratio $\sigma^\mathrm{broad}/\sigma^\mathrm{exp}$ in figure~\ref{fig:multiple_NC_elastic} as a function of the recoil electron energy $E_e^\prime$ in the rest frame of the initial proton for the electron beam energy of the future EicC. We take the same nonperturbative inputs as in subsection~\ref{subsec:one_NC_elastic}. Broadening of electron trajectories inside nuclei reduces the unpolarized cross section by up to a few percent at forward scattering, corresponding to the largest recoil electron energies. This effect decreases at large scattering angles. Cross-section modifications induced by multiple rescattering grow with nuclear size.

\subsection{Charged-current elastic scattering} \label{subsec:multiple_CC_elastic}

\begin{figure}[htb!]
	\centering
	{\includegraphics[angle=0,scale=0.42]{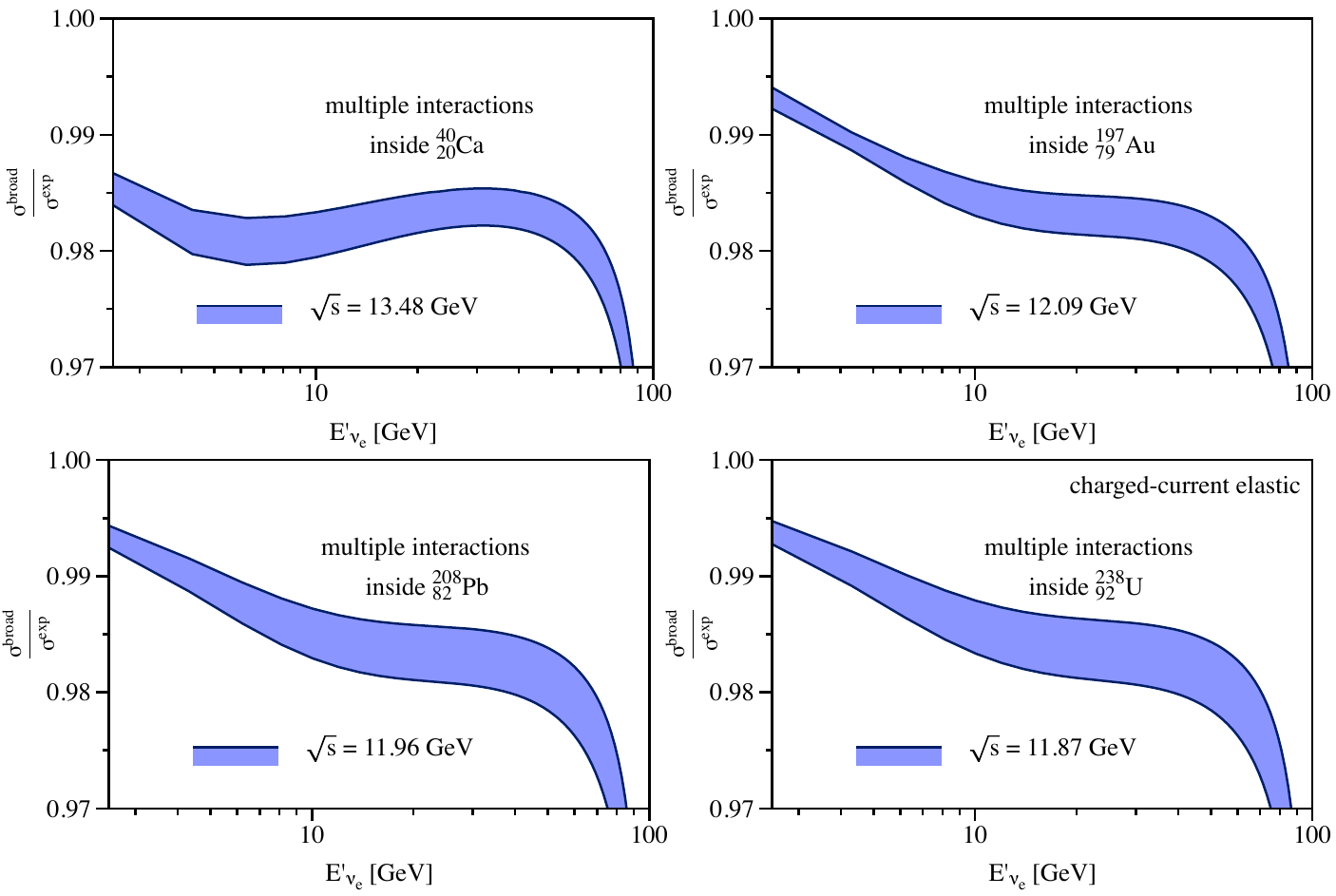}}
	\caption{Ratio of the charged-current elastic electron-proton scattering cross section, after accounting for QED nuclear medium broadening of incoming electrons inside $^{40}_{20}\mathrm{Ca},~^{197}_{79}\mathrm{Au},~^{208}_{82}\mathrm{Pb},$ and $^{238}_{92}\mathrm{U}$ nuclei, to the cross section evaluated from the anticipated lepton kinematics is shown as a function of the recoil neutrino energy $E_{\nu_e}^\prime$ in the rest frame of the initial proton for the electron beam energy of the future EicC. The lower and upper curves correspond to the choice of the atomic scale $\zeta =\frac{m_e Z^{\frac{1}{3}}}{192}$ and $\zeta =\frac{n_\mathrm{max}^2 m_e Z^{\frac{1}{3}}}{192}$, with the smallest and largest principal quantum numbers $1$ and $n_\mathrm{max}$, respectively.}
	\label{fig:multiple_CC_elastic}
\end{figure}
We average the unpolarized charged-current elastic electron-proton scattering cross sections over the transverse momentum distribution of the incoming electron for all possible kinematic configurations of the hard interaction process inside $^{40}_{20}\mathrm{Ca},~^{197}_{79}\mathrm{Au},~^{208}_{82}\mathrm{Pb},$ and $^{238}_{92}\mathrm{U}$ nuclei and present the ratio $\sigma^\mathrm{broad}/\sigma^\mathrm{exp}$ in figure~\ref{fig:multiple_CC_elastic} as a function of the recoil neutrino energy $E_{\nu_e}^\prime$ in the rest frame of the initial proton for the electron beam energy of the future EicC. We take the same nonperturbative inputs as in subsection~\ref{subsec:one_CC_elastic}. As in neutral-current scattering, the correction increases with the recoil lepton energy up to a few percent. In contrast to neutral-current scattering, the correction in charged-current scattering remains finite at backward scattering angles and exhibits a local minimum as a function of the recoil lepton energy, which is more pronounced for rescattering inside lighter nuclei. The dependence of multiple rescattering effects on nuclear size in charged-current reactions is nonmonotonic.

\subsection{Neutral-current inclusive deep inelastic scattering} \label{subsec:multiple_NC_inclusive_DIS}

\begin{figure}[htb!]
	\centering
	{\includegraphics[angle=0,scale=0.378]{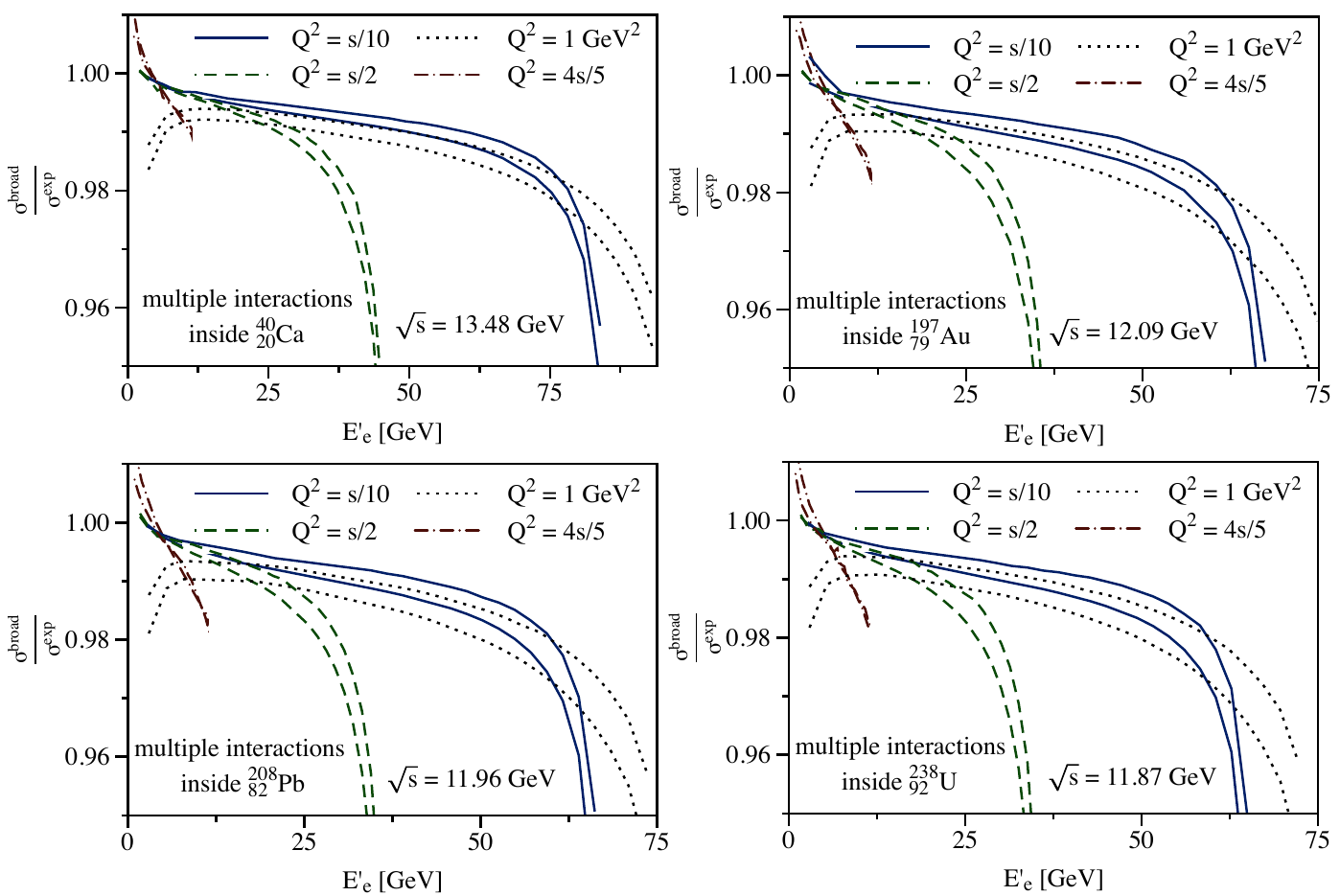}}
	\caption{Ratio of the unpolarized neutral-current inclusive deep inelastic electron-proton scattering cross section, after accounting for QED nuclear medium broadening of incoming and outgoing electrons inside $^{40}_{20}\mathrm{Ca},~^{197}_{79}\mathrm{Au},~^{208}_{82}\mathrm{Pb},$ and $^{238}_{92}\mathrm{U}$ nuclei, to the cross section evaluated from the observed lepton kinematics is shown as a function of the recoil electron energy $E_e^\prime$ in the rest frame of the initial proton for the electron beam energy of the future EicC and fixed values of the squared momentum transfer $Q^2 = 1~\mathrm{GeV}^2,~s/10,~s/2,$ and $4s/5$. The lower and upper curves correspond to the choice of the atomic scale $\zeta =\frac{m_e Z^{\frac{1}{3}}}{192}$ and $\zeta =\frac{n_\mathrm{max}^2 m_e Z^{\frac{1}{3}}}{192}$, with the smallest and largest principal quantum numbers $1$ and $n_\mathrm{max}$, respectively.}
	\label{fig:multiple_NC_inclusive_DIS_x_unpolarized}
	\centering
	{\includegraphics[angle=0,scale=0.378]{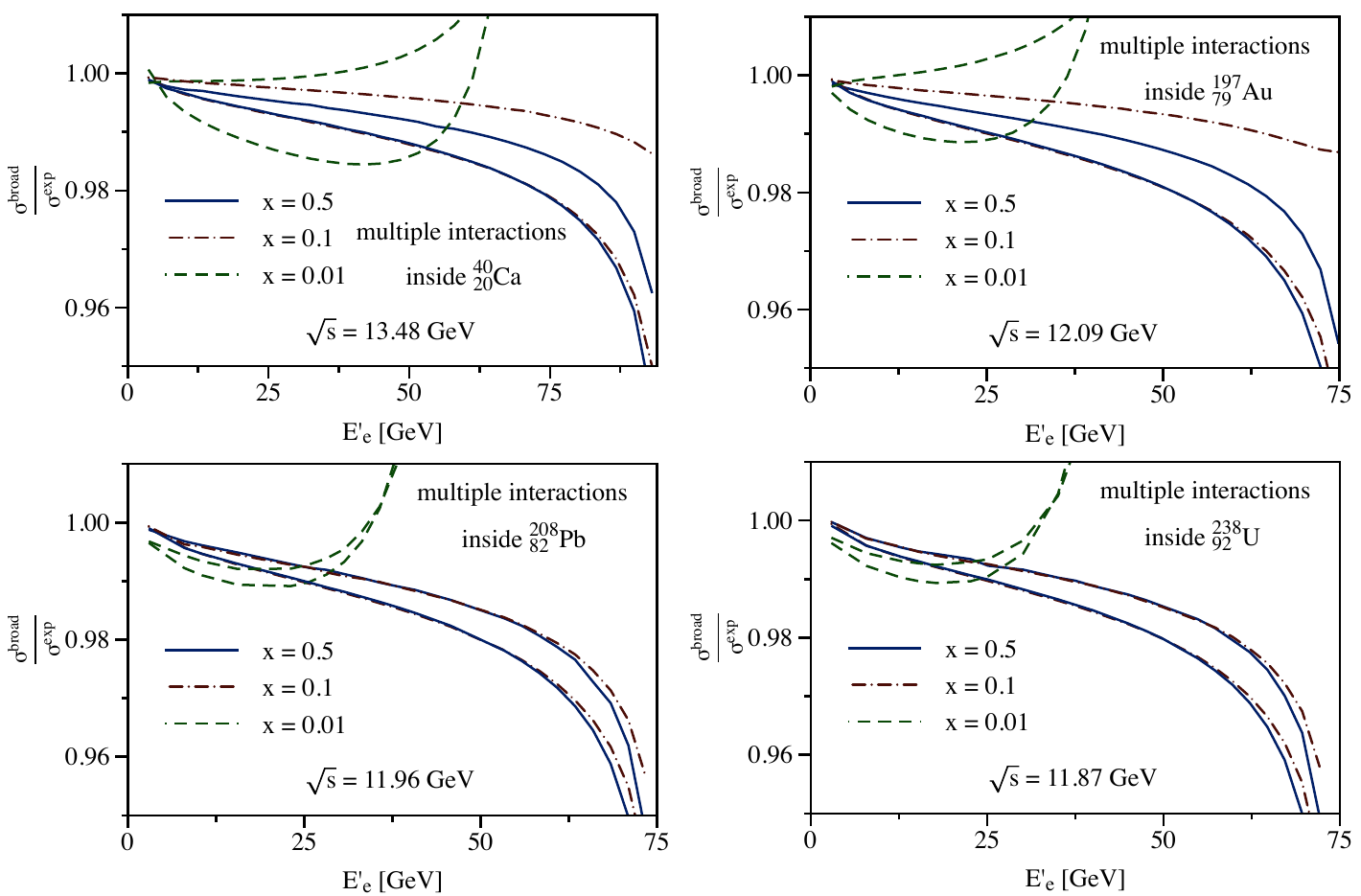}}
	\caption{Ratio of the unpolarized neutral-current inclusive deep inelastic electron-proton scattering cross section, after accounting for QED nuclear medium broadening of incoming and outgoing electrons inside $^{40}_{20}\mathrm{Ca},~^{197}_{79}\mathrm{Au},~^{208}_{82}\mathrm{Pb},$ and $^{238}_{92}\mathrm{U}$ nuclei, to the cross section evaluated from the observed lepton kinematics is shown as a function of the recoil electron energy $E_e^\prime$ in the rest frame of the initial proton for the electron beam energy of the future EicC and fixed values of the Bjorken variable $x = 0.01,~0.1,$ and $0.5$. The lower and upper curves correspond to the choice of the atomic scale $\zeta =\frac{m_e Z^{\frac{1}{3}}}{192}$ and $\zeta =\frac{n_\mathrm{max}^2 m_e Z^{\frac{1}{3}}}{192}$, with the smallest and largest principal quantum numbers $1$ and $n_\mathrm{max}$, respectively.}
	\label{fig:multiple_NC_inclusive_DIS_Q2_unpolarized}
\end{figure}
\newpage
\begin{figure}[htb!]
	\centering
	{\includegraphics[angle=0,scale=0.39]{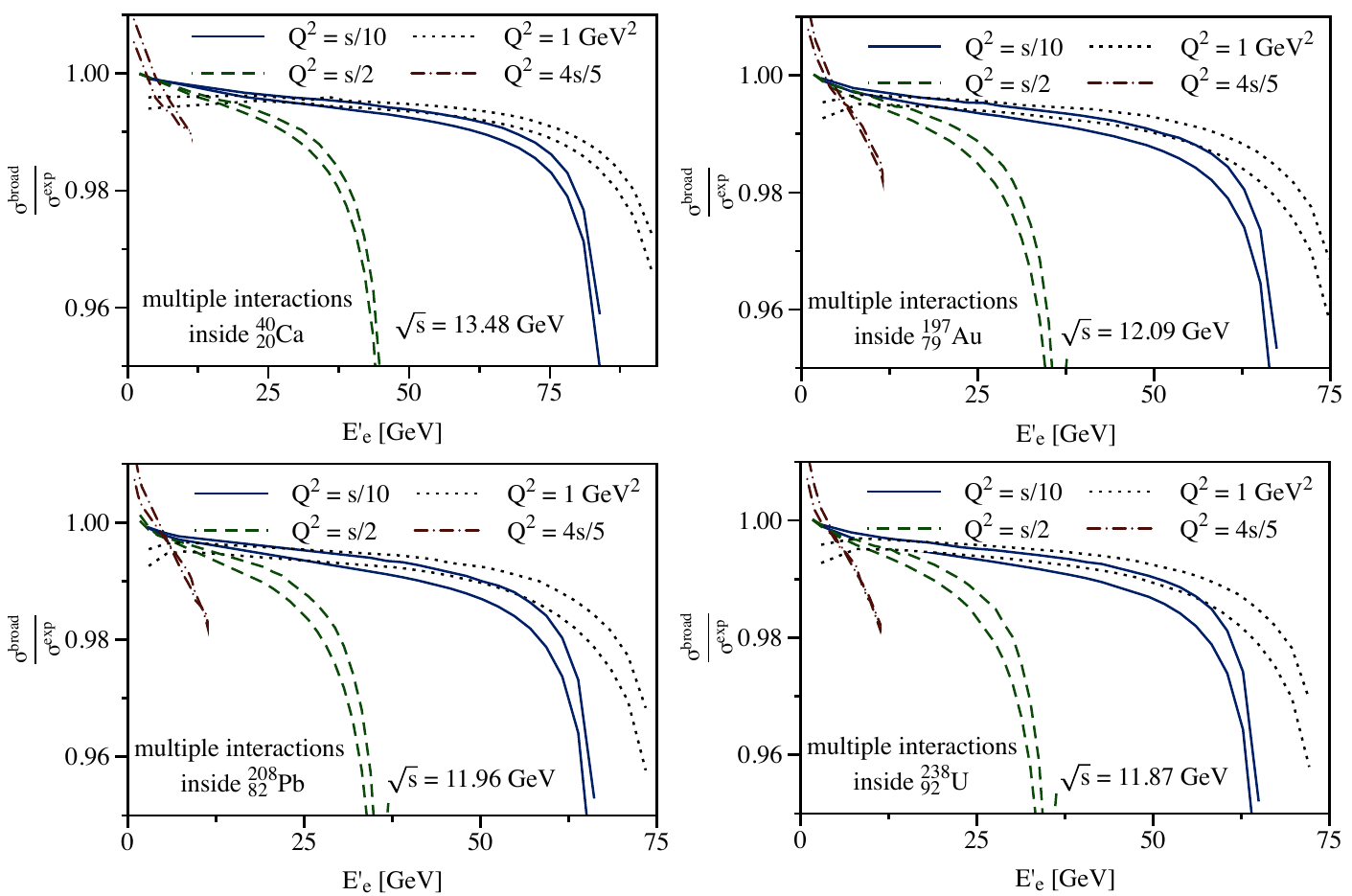}}
	\caption{Ratio of the polarized neutral-current inclusive deep inelastic electron-proton scattering cross section, after accounting for QED nuclear medium broadening of incoming and outgoing electrons inside $^{40}_{20}\mathrm{Ca},~^{197}_{79}\mathrm{Au},~^{208}_{82}\mathrm{Pb},$ and $^{238}_{92}\mathrm{U}$ nuclei, to the cross section evaluated from the observed lepton kinematics is shown as a function of the recoil electron energy $E_e^\prime$ in the rest frame of the initial proton for the electron beam energy of the future EicC and fixed values of the squared momentum transfer $Q^2 = 1~\mathrm{GeV}^2,~s/10,~s/2,$ and $4s/5$. The cross section represents the single-spin asymmetry for longitudinally polarized electrons. The lower and upper curves correspond to the choice of the atomic scale $\zeta =\frac{m_e Z^{\frac{1}{3}}}{192}$ and $\zeta =\frac{n_\mathrm{max}^2 m_e Z^{\frac{1}{3}}}{192}$, with the smallest and largest principal quantum numbers $1$ and $n_\mathrm{max}$, respectively.}
	\label{fig:multiple_NC_inclusive_DIS_x_polarized}
	\centering
	{\includegraphics[angle=0,scale=0.39]{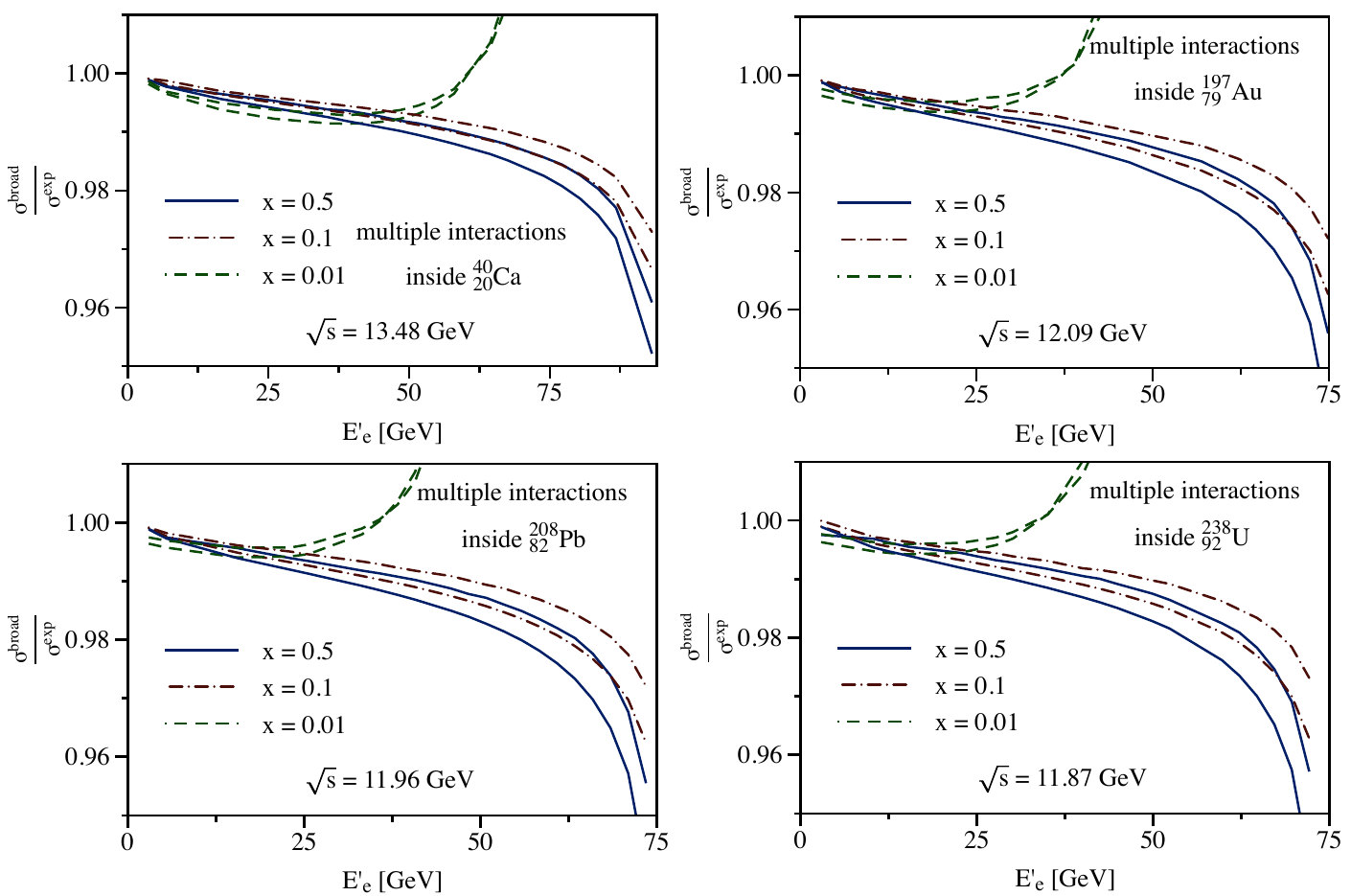}}
	\caption{Ratio of the polarized neutral-current inclusive deep inelastic electron-proton scattering cross section, after accounting for QED nuclear medium broadening of incoming and outgoing electrons inside $^{40}_{20}\mathrm{Ca},~^{197}_{79}\mathrm{Au},~^{208}_{82}\mathrm{Pb},$ and $^{238}_{92}\mathrm{U}$ nuclei, to the cross section evaluated from the observed lepton kinematics is shown as a function of the recoil electron energy $E_e^\prime$ in the rest frame of the initial proton for the electron beam energy of the future EicC and fixed values of the Bjorken variable $x = 0.01,~0.1,$ and $0.5$. The cross section represents the single-spin asymmetry for longitudinally polarized electrons. The lower and upper curves correspond to the choice of the atomic scale $\zeta =\frac{m_e Z^{\frac{1}{3}}}{192}$ and $\zeta =\frac{n_\mathrm{max}^2 m_e Z^{\frac{1}{3}}}{192}$, with the smallest and largest principal quantum numbers $1$ and $n_\mathrm{max}$, respectively.}
	\label{fig:multiple_NC_inclusive_DIS_Q2_polarized}
\end{figure}
We average the unpolarized neutral-current inclusive deep inelastic electron-proton scattering cross sections over the transverse momentum distributions of the initial and final electrons for all possible kinematic configurations of the hard interaction process inside $^{40}_{20}\mathrm{Ca},~^{197}_{79}\mathrm{Au},~^{208}_{82}\mathrm{Pb},$ and $^{238}_{92}\mathrm{U}$ nuclei and present the ratio $\sigma^\mathrm{broad}/\sigma^\mathrm{exp}$ as a function of the recoil electron energy $E_e^\prime$ in the rest frame of the initial proton for the electron beam energy of the future EicC. We take the same nonperturbative inputs as in subsection~\ref{subsec:one_NC_inclusive_DIS}.

For fixed values of the squared momentum transfer $Q^2 = 1~\mathrm{GeV}^2,~s/10,~s/2,$ and $4s/5$, the results for unpolarized inclusive DIS off the proton are shown in figure~\ref{fig:multiple_NC_inclusive_DIS_x_unpolarized}, within the allowed kinematic ranges. QED nuclear medium effects increase with the recoil electron energy, reaching the few-percent level and exhibiting a nonmonotonic dependence on nuclear size. We present the results for unpolarized inclusive DIS off the proton at fixed values of the Bjorken variable $x = 0.01,~0.1,$ and $0.5$ in figure~\ref{fig:multiple_NC_inclusive_DIS_Q2_unpolarized}. The effects are larger in scattering at small angles and at small Bjorken variable $x$. At large $x$, multiple rescattering corrections increase with the nuclear size.

The ratio of the ``true" over the expected cross section $\sigma^\mathrm{broad}/\sigma^\mathrm{exp}$ after accounting for QED nuclear medium rescattering of incoming and outgoing electrons in polarized inclusive electron-proton DIS is shown at fixed values of the squared momentum transfer $Q^2 = 1~\mathrm{GeV}^2,~s/10,~s/2,$ and $4s/5$ in figure~\ref{fig:multiple_NC_inclusive_DIS_x_polarized} and at fixed values of the Bjorken variable $x = 0.01,~0.1,$ and $0.5$ in figure~\ref{fig:multiple_NC_inclusive_DIS_Q2_polarized}. The qualitative behavior of QED nuclear medium effects is the same as for unpolarized scattering, with a smaller relative cross-section correction for polarized contributions to inclusive electron-proton DIS.

\subsection{Charged-current inclusive deep inelastic scattering} \label{subsec:multiple_CC_inclusive_DIS}

\begin{figure}[htb!]
	\centering
	{\includegraphics[angle=0,scale=0.39]{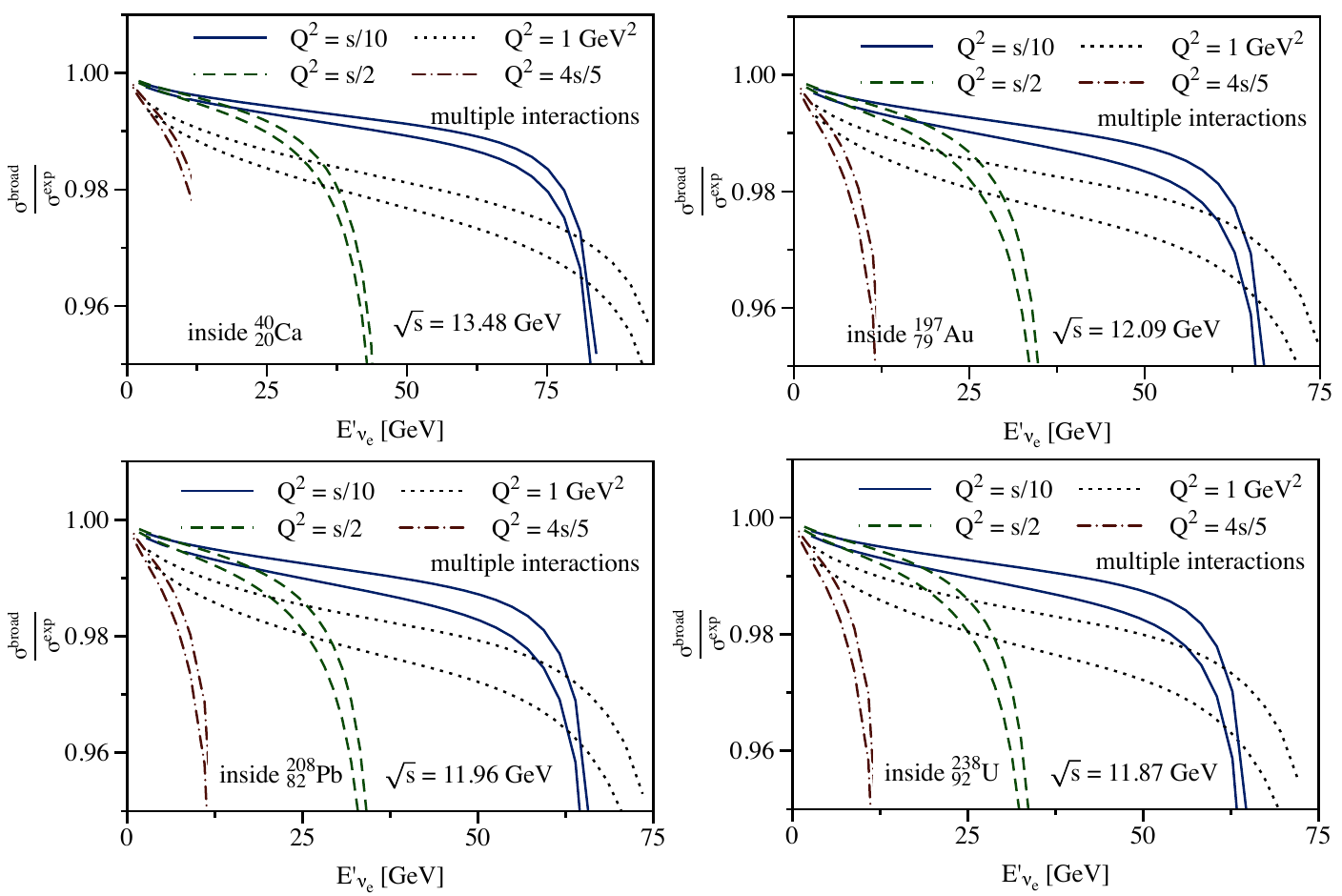}}
	\caption{Ratio of the unpolarized charged-current inclusive deep inelastic electron-proton scattering cross section, after accounting for QED nuclear medium broadening of incoming electrons inside $^{40}_{20}\mathrm{Ca},~^{197}_{79}\mathrm{Au},~^{208}_{82}\mathrm{Pb},$ and $^{238}_{92}\mathrm{U}$ nuclei, to the cross section evaluated from the anticipated lepton kinematics is shown as a function of the recoil neutrino energy $E_{\nu_e}^\prime$ in the rest frame of the initial proton for the electron beam energy of the future EicC and fixed values of the squared momentum transfer $Q^2 = 1~\mathrm{GeV}^2,~s/10,~s/2,$ and $4s/5$. The lower and upper curves correspond to the choice of the atomic scale $\zeta =\frac{m_e Z^{\frac{1}{3}}}{192}$ and $\zeta =\frac{n_\mathrm{max}^2 m_e Z^{\frac{1}{3}}}{192}$, with the smallest and largest principal quantum numbers $1$ and $n_\mathrm{max}$, respectively.}
	\label{fig:multiple_CC_inclusive_DIS_x_unpolarized}
	\centering
	{\includegraphics[angle=0,scale=0.39]{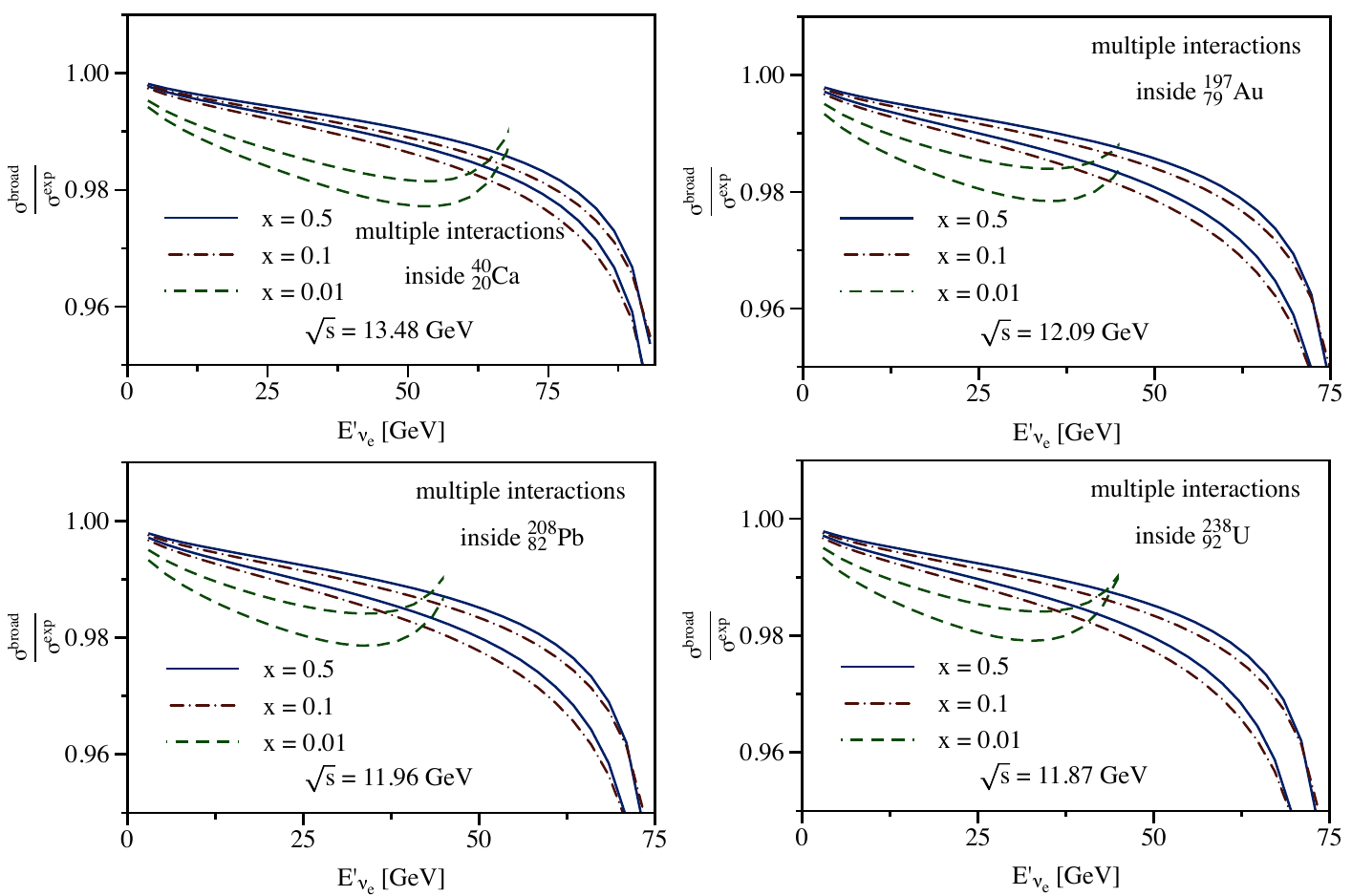}}
	\caption{Ratio of the unpolarized charged-current inclusive deep inelastic electron-proton scattering cross section, after accounting for QED nuclear medium broadening of incoming electrons inside $^{40}_{20}\mathrm{Ca},~^{197}_{79}\mathrm{Au},~^{208}_{82}\mathrm{Pb},$ and $^{238}_{92}\mathrm{U}$ nuclei, to the cross section evaluated from the anticipated lepton kinematics is shown as a function of the recoil neutrino energy $E_{\nu_e}^\prime$ in the rest frame of the initial proton for the electron beam energy of the future EicC and fixed values of the Bjorken variable $x = 0.01,~0.1,$ and $0.5$. The lower and upper curves correspond to the choice of the atomic scale $\zeta =\frac{m_e Z^{\frac{1}{3}}}{192}$ and $\zeta =\frac{n_\mathrm{max}^2 m_e Z^{\frac{1}{3}}}{192}$, with the smallest and largest principal quantum numbers $1$ and $n_\mathrm{max}$, respectively.}
	\label{fig:multiple_CC_inclusive_DIS_Q2_unpolarized}
\end{figure}
We average the unpolarized charged-current inclusive deep inelastic electron-proton scattering cross sections over the transverse momentum distributions of the incoming electron for all possible kinematic configurations of the hard interaction process inside $^{40}_{20}\mathrm{Ca},$ $^{197}_{79}\mathrm{Au},~^{208}_{82}\mathrm{Pb},$ and $^{238}_{92}\mathrm{U}$ nuclei and present the ratio $\sigma^\mathrm{broad}/\sigma^\mathrm{exp}$ as a function of the recoil (anti)neutrino energy $E_{\nu_e}^\prime$ in the rest frame of the initial proton for the electron beam energy of the future EicC. We take the same nonperturbative inputs as in subsection~\ref{subsec:one_CC_inclusive_DIS}.

Results for the ratio $\sigma^\mathrm{broad}/\sigma^\mathrm{exp}$ of the ``true" to the expected cross section after accounting for QED nuclear medium rescattering are shown in figure~\ref{fig:multiple_CC_inclusive_DIS_x_unpolarized} at fixed values of the squared momentum transfer $Q^2 = 1~\mathrm{GeV}^2,~s/10,~s/2,$ and $4s/5$, and in figure~\ref{fig:multiple_CC_inclusive_DIS_Q2_unpolarized} at fixed values of the Bjorken variable $x = 0.01,~0.1,$ and $0.5$. QED nuclear medium effects increase with the recoil neutrino energy, reaching the few-percent level and exhibiting a nonmonotonic dependence on nuclear size. The effects are the largest in forward scattering and at small Bjorken variable $x$. Qualitatively, the results for charged-current scattering are the same as for neutral-current scattering. Quantitatively, corrections in charged-current inclusive DIS are smaller than in unpolarized neutral-current inclusive DIS and comparable to effects in polarized neutral-current inclusive DIS.

\section{Conclusions and Outlook} \label{sec:conclusions_and_outlook}

In this paper, we present evaluations of QED nuclear medium effects inside the anticipated $^{197}_{79}\mathrm{Au},~^{208}_{82}\mathrm{Pb},$ and $^{238}_{92}\mathrm{U}$ nuclei at the future Electron-ion collider in China for the expected beam energy. We also extrapolate these calculations to medium-size nuclei and present the results for the $^{40}_{20}\mathrm{Ca}$ nucleus. We study neutral-current and charged-current, unpolarized and polarized, elastic and inclusive deep inelastic electron scattering inside large nuclei. For charged-current electron scattering and polarized contributions to inclusive deep inelastic electron scattering, we provide the first calculations.

At first order in the opacity expansion, the QED nuclear medium cross-section modifications are larger at small scattering angles. The corrections reach the percent level in elastic scattering. They are positive-definite in neutral-current elastic scattering and change sign in charged-current elastic scattering as a function of kinematics, taking negative values at small scattering angles.

QED nuclear medium corrections at first order in the opacity expansion and effects from kinematic distortions after multiple soft rescattering in charged-current inclusive DIS are smaller than those in unpolarized neutral-current inclusive DIS and comparable to effects in polarized neutral-current inclusive DIS. The qualitative behavior in all these cases is the same, with corrections generally increasing toward small scattering angles and large $x$, except for the monotonic dependence on $x$ in charged-current inclusive DIS. In contrast to neutral-current elastic scattering, the correction in charged-current elastic scattering remains finite at backward scattering angles and exhibits a local minimum as a function of the recoil lepton energy, a feature that is more pronounced for lighter nuclei.

QED nuclear medium corrections at first order in the opacity expansion and effects from kinematic distortions after multiple soft rescattering from sections~\ref{sec:one} and~\ref{sec:multiple}, respectively, have similar magnitudes, cannot be distinguished experimentally, and should be added incoherently. This results in a numerical suppression of the total correction at forward angles, with multiple rescattering becoming the dominant mechanism in nonforward kinematics.

\appendix

\section{Charged-current elastic electron-nucleon scattering cross section} \label{app:elastic_xsec}

In this appendix, we describe the cross section for charged-current elastic electron-proton scattering,
\begin{align}
	e^- \left( p \right) p \left( k \right) &\to \nu_e \left( p^\prime \right) n \left( k^\prime \right), \label{eq:neutrino_processes}
\end{align} 
at leading order in the electromagnetic coupling constant for vanishing electron mass.

The unpolarized differential cross section is conveniently expressed in terms of the structure-dependent $A,~B$, and $C$ parameters as~\cite{LlewellynSmith:1971uhs,Yang:2026vuf}
\begin{equation}
	\frac{d\sigma_\nu}{dQ^2} (Q^2, E_e) = \frac{\mathrm{G}_\mathrm{F}^2 |V_{u d}|^2}{2 \pi} \frac{M^2}{E_e^2} \left[ \tau A(Q^2) - \nu B(Q^2) + \frac{\nu^2}{1+\tau} C(Q^2) \right] \,, \label{eq:xsection_neutrino}
\end{equation}
where $\tau = Q^2 / \left( 4 M^2 \right)$, $\nu = E_e/M - \tau$, $\mathrm{G}_\mathrm{F}$ is the Fermi coupling constant, and $V_{u d}$ is the Cabibbo-Kobayashi-Maskawa matrix element. Assuming isospin symmetry, the structure-dependent factors $A,~B$, and $C$ are expressed in terms of the isovector-vector Sachs electric, $G^V_E$, and magnetic $G^V_M$, axial-vector, $F_A$, and pseudoscalar, $F_P$, form factors, which are functions of $Q^2$ only, as
\begin{align}
	A &= \tau \left( G^V_M \right)^2 - \left( G^V_E \right)^2 + (1+ \tau) F_A^2\,, \\
	B &= 4 \tau F_A G^V_M \,, \\
	C &= \tau \left( G^V_M \right)^2 + \left( G^V_E \right)^2 + (1+ \tau) F_A^2 \,. \label{eq:ABC}
\end{align}

In the limit of isospin symmetry, both the electric and magnetic isovector-vector form factors are given by the difference of the proton and neutron form factors, i.e., $G_{E,M}^V = G_{E,M}^p-G_{E,M}^n$. $Q^2$ ranges from $0$ in the forward direction to $Q^2_\mathrm{+}$ in backward scattering,
\begin{equation}
	Q^2_\mathrm{+} = \frac{4 M E^2_\nu}{M+2 E_\nu}. \label{eq:charged_current_Q2_range}
\end{equation}

The contribution from left-handed electron polarization completely determines the unpolarized component, while right-handed electrons do not contribute to the cross section in the limit of vanishing electron mass. That is why QED nuclear medium effects in unpolarized and polarized charged-current scattering are the same up to electron-mass corrections.

\section{Inclusive electron-nucleon DIS cross sections} \label{app:inclusive_xsec}

In this appendix, we provide expressions for neutral-current and charged-current inclusive deep inelastic electron-nucleon scattering cross sections in terms of the nucleon structure functions and quark parton distribution functions. The unpolarized cross section and cross section with a polarized electron are presented.

The inclusive deep inelastic electron-nucleon scattering cross section, in the approximation of one exchanged gauge boson and vanishing electron mass, is traditionally expressed as a product of the leptonic $L_i^{\mu \nu}$ and hadronic $W^i_{\mu \nu}$ tensors~\cite{Halzen:1984mc,CTEQ:1993hwr,ParticleDataGroup:2024cfk}:\footnote{The electromagnetic coupling constant in the hard scattering process away from the Thomson limit cancels in all ratios in this paper. The running of $\sin^2 \theta_W$, which results in a relative uncertainty below $4\%$, is neglected.}
\begin{align}
	\mathrm{d} \sigma = \frac{1}{2 \left( s - M^2 \right)}\frac{\left(4 \pi \alpha \right)^2}{Q^4} \left( 4 \pi \sum \limits_{i = \gamma, \gamma Z, Z; W} \eta_i L_i^{\mu \nu} W^i_{\mu \nu} \right) \frac{\mathrm{d}^3 p^\prime}{\left( 2 \pi \right)^3 2 E^\prime},
\end{align}
with the factors $\eta_i$ that depend on the exchanged gauge boson $i$ as
\begin{equation}
	\eta_\gamma = 1, \qquad \eta_{\gamma Z} = \frac{1}{\sin^2 \left( 2 \theta_W \right) } \frac{Q^2}{Q^2+M_Z^2}, \qquad \eta_Z = \eta_{\gamma Z}^2, \qquad \eta_W = \frac{\eta_{\gamma Z}^2}{4} \frac{\left( Q^2+M_Z^2 \right)^2}{\left( Q^2+M_W^2 \right)^2} ,
\end{equation}
with the mass of the $Z$ boson $M_Z$, the mass of the $W$ boson $M_W$, and the Weinberg angle $\theta_W$.\footnote{We take the $\overline{\mathrm{MS}}$ values for $\sin^2 \theta_W$ and $M_Z$ at the scale $\mu = M_Z$ from Ref.~\cite{Hill:2019xqk}.} The explicit expressions for the tensors $L_i^{\mu \nu}$ and $W^i_{\mu \nu}$ are
\begin{align}
	L_\gamma^{\mu \nu} &= 2 \left( p^\mu \left(p^\prime\right)^\nu + \left(p^\prime \right)^\mu p^\nu - g^{\mu \nu} \frac{Q^2}{2} - i \lambda^e \varepsilon^{\mu \nu \alpha \beta} p_\alpha p^\prime_\beta \right), \\[2ex]
	L_{\gamma Z}^{\mu \nu} &= - Q_e \left( g_V^e + \lambda^e g_A^e \right) L_\gamma^{\mu \nu}, \\[2ex]
	L_Z^{\mu \nu} &= \left( g_V^e + \lambda^e g_A^e \right)^2 L_\gamma^{\mu \nu}, \\[2ex]
	L_W^{\mu \nu} &= \left( 1 + \lambda_e \right)^2 L_\gamma^{\mu \nu},
\end{align}
\begin{align}
	W^\gamma_{\mu \nu} &= \frac{1}{8 \pi} \sum \limits_{S, q, X} <N \left( k, S \right) | \left[ J_\mu^{q,\gamma} \right]^\dagger| X > < X | J_\nu^{q,\gamma}| N \left( k, S \right) > \left( 2 \pi \right)^4 \delta^{4} \left( k + q - p_X \right), \\[2ex]
	W^{\gamma Z}_{\mu \nu} &= \frac{1}{8 \pi} \sum \limits_{S, q, X} <N \left( k, S \right) | \left[ J_\mu^{q,Z} \right]^\dagger | X > < X | J_\nu^{q,\gamma} | N \left( k, S \right) > \left( 2 \pi \right)^4 \delta^{4} \left( k + q - p_X \right) \nonumber \\
    &+\frac{1}{8 \pi} \sum \limits_{S, q, X} <N \left( k, S \right) | \left[ J_\mu^{q,\gamma} \right]^\dagger | X > < X | J_\nu^{q,Z} | N \left( k, S \right) > \left( 2 \pi \right)^4 \delta^{4} \left( k + q - p_X \right), \\[2ex]
	W^Z_{\mu \nu} &= \frac{1}{8 \pi} \sum \limits_{S, q, X} <N \left( k, S \right) | \left[ J_\mu^{q,Z} \right]^\dagger | X > < X | J_\nu^{q,Z} | N \left( k, S \right) > \left( 2 \pi \right)^4 \delta^{4} \left( k + q - p_X \right), \\[2ex]
	W^W_{\mu \nu} &= \frac{1}{8 \pi} \sum \limits_{S, q, X} <N \left( k, S \right) | \left[ J_\mu^{q,W} \right]^\dagger | X > < X | J_\nu^{q,W} | N \left( k, S \right) > \left( 2 \pi \right)^4 \delta^{4} \left( k + q - p_X \right),
\end{align}
with the electron helicity $\lambda^e = \pm 1$. The sum is performed over all spin states $S$ of the nucleon, all quarks and antiquarks, and all allowed hadronic final states $X$, with the integration over the four-momentum $p_X$. The electromagnetic $J_\mu^{q,\gamma}$, neutral $J_\mu^{q,Z}$, and charged $J_\mu^{q,W}$ quark currents are expressed in terms of the quark fields $q$ as $J_\mu^{q,\gamma} = - Q_q \bar{q} \gamma_\mu q,~J_\mu^{q,Z} = \bar{q} \gamma_\mu \left( g_V^q - g_A^q \gamma_5 \right) q,$ and $J_\mu^{q,W} = \bar{q} \tau^- \gamma_\mu P_L q^\prime$, with the isospin-lowering operator $\tau^-$, respectively. At leading order, the vector $g_V^f$ and axial-vector $g_A^f$ coupling constants of quarks and electrons, where $f$ denotes the fermion, are expressed in terms of the electric charge $Q_f$ and the third component of the isospin $T_f^3$ as
\begin{equation}
	g_V^f = T_f^3 - 2 Q_f \sin^2 \theta_W, \qquad g_A^f =T_f^3.
\end{equation}
The electron electric charge and the third component of the isospin are fixed as $Q_e = -1$ and $T_e^3 = -\frac{1}{2}$, respectively, while the quark quantum numbers are $Q_u = \frac{2}{3},~Q_d = - \frac{1}{3}$ and $T_u^3 = \frac{1}{2},~T_d^3 = - \frac{1}{2} $ for $u$- and $d$-type quarks, respectively.

The Lorentz-invariant decomposition for the part of the hadronic tensor $W^i_{\mu \nu}$ that contributes to the electron-nucleon scattering cross sections of interest can be expressed in terms of the nucleon structure functions $F^i_1$, $F^i_2$, and $F^i_3$ as
\begin{align}
	W^i_{\mu \nu} = &\left( - g_{\mu \nu} + \frac{q_\mu q_\nu}{q^2} \right) F^i_1 \left( x, Q^2 \right) + \left( k_\mu - \frac{k \cdot q}{q^2} q_\mu \right) \left( k_\nu - \frac{k \cdot q}{q^2} q_\nu \right) \frac{F^i_2 \left( x, Q^2 \right)}{k \cdot q} \nonumber \\
	&- i \varepsilon_{ \mu \nu \alpha \beta} q^\alpha \left( k^\beta - \frac{k \cdot q}{q^2} q^\beta \right) \frac{F^i_3 \left( x, Q^2 \right)}{2 k \cdot q}.
\end{align}
At tree level, the nucleon structure functions are written in terms of the parton distribution functions $q \left( x, Q^2 \right)$ of quarks and antiquarks, with an opposite sign in $F_3$ for antiquarks, inside the nucleus as
\begin{align}
	2 x F^\gamma_{1} \left( x, Q^2\right) &= F^\gamma_2 \left( x, Q^2\right) = x \sum \limits_q Q^2_q q \left( x, Q^2 \right), \quad F^\gamma_3 \left( x, Q^2\right) = 0, \\
	2 x F^{\gamma Z}_{1} \left( x, Q^2\right) &= F^{\gamma Z}_2 \left( x, Q^2\right) = 2 x \sum \limits_q Q_q g_V^q q \left( x, Q^2 \right), \\
\ F^{\gamma Z}_3 \left( x, Q^2\right) &= 2 \sum \limits_q Q_q g_A^q q \left( x, Q^2 \right), \\
	2 x F^Z_{1} \left( x, Q^2\right) &= F^Z_2 \left( x, Q^2\right) = x \sum \limits_q \left[ \left(g_V^q \right)^2 + \left(g_A^q \right)^2 \right] q \left( x, Q^2 \right), \\
	F^Z_3 \left( x, Q^2\right) &= 2 \sum \limits_q g_V^q g_A^q q \left( x, Q^2 \right), \\
	2 x F^W_{1} \left( x, Q^2\right) &= F^W_2 \left( x, Q^2\right) = 2 x \sum \limits_q \left[ \left( g_A^q + \frac{1}{2} \right) q \left( x, Q^2 \right) - \left( g_A^q - \frac{1}{2} \right) \overline{q} \left( x, Q^2 \right) \right], \\
	F^W_3 \left( x, Q^2\right) &= 2 \sum \limits_q \left[ \left( g_A^q + \frac{1}{2} \right) q \left( x, Q^2 \right) - \left( g_A^q - \frac{1}{2} \right) \overline{q} \left( x, Q^2 \right) \right].
\end{align}

The double-differential inclusive deep inelastic electron-nucleon scattering cross section is expressed in terms of the nucleon structure functions $F_1$, $F_2$, and $F_3$, and the kinematic invariants as
\begin{align}
	\frac{\mathrm{d}^2 \sigma}{\mathrm{d} x \mathrm{d} Q^2} = \dfrac{4 \pi \alpha^{2}}{x Q^4} \bigg [ x y^2 F_1 (x, Q^2) + \left( 1 - y - \frac{x^2 y^2 M^2}{Q^2} \right)F_2 (x, Q^2) + \left( y - \frac{y^2}{2} \right) x F_3 (x, Q^2) \bigg ],
\end{align}
with the variable $y = \frac{Q^2}{ \left( s - M^2 \right) x}$, which is unaffected by QED nuclear medium effects. The structure functions $F_1$, $F_2$, and $F_3$ for neutral-current and charged-current scattering are expressed as
\begin{align}
	F^\mathrm{NC}_{1,2} &= F^\gamma_{1,2} + Q_e \left( g_V^e + \lambda^e g_A^e \right) \eta_{\gamma Z} F^{\gamma Z}_{1,2} + \left[ \left(g_V^e \right)^2 + \left(g_A^e \right)^2 + 2 \lambda^e g_V^e g_A^e \right] \eta_{Z} F^{Z}_{1,2}, \\
	F^\mathrm{NC}_{3} &= F^\gamma_{3} + Q_e \left( g_A^e + \lambda^e g_V^e\right) \eta_{\gamma Z} F^{\gamma Z}_{3} + \lambda^e \left[ \left(g_V^e \right)^2 + \left(g_A^e \right)^2 + 2 \lambda^e g_V^e g_A^e \right] \eta_{Z} F^{Z}_{3}, \\
	F^\mathrm{CC}_{1,2} &= \eta_{W}\left(1 + \lambda^e \right)^2 F^{W}_{1,2}, \\
	F^\mathrm{CC}_{3} &= \eta_{W} \left(1 + \lambda^e \right)^2 F^{W}_{3}.
\end{align}
Positron scattering can be obtained from electron scattering by flipping the sign of the lepton helicity $\lambda^e$ and the structure function $F_3$.

As in charged-current elastic scattering, QED nuclear medium effects in polarized charged-current inclusive DIS are the same as in unpolarized charged-current inclusive DIS up to electron-mass corrections.

\acknowledgments

The work of O.T. is supported by the National Science Foundation of China under Grants No. 12347105 and No. 12447101. FeynCalc~\cite{Mertig:1990an,Shtabovenko:2016sxi}, Mathematica~\cite{Mathematica}, and DataGraph~\cite{JSSv047s02} were used in this work.

\bibliographystyle{JHEP}
\bibliography{references}

@article{Xiao:2026tbs,
    author = "Xiao, Bo-Wen and Zhao, Yuxiang and Zhou, Jian",
    title = "{Physics of the Electron-Ion Collider in China}",
    eprint = "2608.11712",
    archivePrefix = "arXiv",
    primaryClass = "hep-ph",
    month = "8",
    year = "2026"
}

@article{Kuraev:2013sea,
    author = "Kuraev, E. A. and Voskresenskaya, O. O. and Torosyan, H. T.",
    title = "{Coulomb corrections to the parameters of the Landau-Pomeranchuk-Migdal effect theory}",
    eprint = "1309.7946",
    archivePrefix = "arXiv",
    primaryClass = "hep-ph",
    doi = "10.1134/S1547477114040281",
    journal = "Phys. Part. Nucl. Lett.",
    volume = "11",
    pages = "366--380",
    year = "2014"
}

@techreport{JLab-PR12-25-009,
  author       = {Averett, T. and Napolitano, J. and Wojtsekhowski, B. and Xiong, W., et. al},
  title        = {The Nucleon Axial-Vector Form Factor from the H($\vec{e}$, n)$\nu_e$ Reaction},
  institution  = {Thomas Jefferson National Accelerator Facility},
  address      = {Newport News, VA},
  number       = {PR12-25-009},
  type         = {Proposal},
  year         = {2025},
  month        = {July},
  note         = {Proposal to JLab PAC 53},
  url          = {https://indico.jlab.org/event/923/contributions/17312/}
}

@article{Klest:2025bfl,
    author = "Klest, Henry T.",
    title = "{Charged-current elastic scattering at the Electron-Ion Collider}",
    eprint = "2511.02049",
    archivePrefix = "arXiv",
    primaryClass = "nucl-ex",
    doi = "10.1103/1cgs-6ksy",
    journal = "Phys. Rev. D",
    volume = "113",
    number = "3",
    pages = "033002",
    year = "2026"
}

@article{Yang:2026vuf,
    author = "Yang, Guang and Kumar, Praveen",
    title = "{Constraining neutrino-nucleon form factors with charged-current scattering at the Electron-Ion Collider}",
    eprint = "2603.00703",
    archivePrefix = "arXiv",
    primaryClass = "hep-ph",
    doi = "10.1103/btj9-n88g",
    journal = "Phys. Rev. D",
    volume = "113",
    number = "11",
    pages = "116031",
    year = "2026"
}

@article{Davoudiasl:2025ifk,
    author = "Davoudiasl, Hooman and Liu, Hongkai and Mantry, Sonny and Neil, Ethan T.",
    title = "{Weak-charge form-factor determination at the electron-ion collider}",
    eprint = "2512.15865",
    archivePrefix = "arXiv",
    primaryClass = "hep-ph",
    doi = "10.1103/lnqv-bltq",
    journal = "Phys. Rev. D",
    volume = "113",
    number = "5",
    pages = "L051301",
    year = "2026"
}

@article{Fatima:2026mac,
    author = "Fatima, A. and Sajjad Athar, M. and Singh, S. K.",
    title = "{Charged current induced electron-proton scattering and the axial vector form factor}",
    eprint = "2604.00764",
    archivePrefix = "arXiv",
    primaryClass = "hep-ph",
    doi = "10.1103/bffn-fzgx",
    journal = "Phys. Rev. D",
    volume = "114",
    number = "1",
    pages = "013004",
    year = "2026"
}

@article{Fatima:2026hyc,
    author = "Fatima, A. and Sajjad Athar, M. and Singh, S. K.",
    title = "{Weak charged current induced electron and positron scattering off proton at JLab and MAMI energies}",
    eprint = "2607.13523",
    archivePrefix = "arXiv",
    primaryClass = "hep-ph",
    month = "7",
    year = "2026"
}

@article{Campbell:2022qmc,
    author = "Campbell, J. M. and others",
    title = "{Event generators for high-energy physics experiments}",
    eprint = "2203.11110",
    archivePrefix = "arXiv",
    primaryClass = "hep-ph",
    reportNumber = "CP3-22-12, DESY-22-042, FERMILAB-PUB-22-116-SCD-T, IPPP/21/51,
  JLAB-PHY-22-3576, KA-TP-04-2022, LA-UR-22-22126, LU-TP-22-12, MCNET-22-04,
  OUTP-22-03P, P3H-22-024, PITT-PACC 2207, UCI-TR-2022-02",
    doi = "10.21468/SciPostPhys.16.5.130",
    journal = "SciPost Phys.",
    volume = "16",
    number = "5",
    pages = "130",
    year = "2024"
}

@article{TenaVidal:2024eyt,
    author = "Tena Vidal, Julia and Ashkenazi, Adi and Weinstein, L. B. and Blunden, Peter and Dytman, Steven and Steinberg, Noah",
    title = "{A universal implementation of radiative effects in neutrino event generators}",
    eprint = "2409.05736",
    archivePrefix = "arXiv",
    primaryClass = "hep-ex",
    reportNumber = "FERMILAB-PUB-24-0592-PPD",
    doi = "10.1016/j.cpc.2025.109509",
    journal = "Comput. Phys. Commun.",
    volume = "310",
    pages = "109509",
    year = "2025"
}

@article{Banerjee:2020rww,
    author = "Banerjee, Pulak and Engel, T. and Signer, A. and Ulrich, Y.",
    title = "{QED at NNLO with McMule}",
    eprint = "2007.01654",
    archivePrefix = "arXiv",
    primaryClass = "hep-ph",
    reportNumber = "PSI-PR-20-09, ZU-TH 23/20",
    doi = "10.21468/SciPostPhys.9.2.027",
    journal = "SciPost Phys.",
    volume = "9",
    pages = "027",
    year = "2020"
}

@article{Hofstadter:1956qs,
    author = "Hofstadter, Robert",
    title = "{Electron scattering and nuclear structure}",
    doi = "10.1103/RevModPhys.28.214",
    journal = "Rev. Mod. Phys.",
    volume = "28",
    pages = "214--254",
    year = "1956"
}

@article{Ernst:1960zza,
    author = "Ernst, F. J. and Sachs, R. G. and Wali, K. C.",
    title = "{Electromagnetic form factors of the nucleon}",
    doi = "10.1103/PhysRev.119.1105",
    journal = "Phys. Rev.",
    volume = "119",
    pages = "1105--1114",
    year = "1960"
}

@article{Hand:1963zz,
    author = "Hand, L. N. and Miller, D. G. and Wilson, Richard",
    title = "{Electric and Magnetic Formfactor of the Nucleon}",
    doi = "10.1103/RevModPhys.35.335",
    journal = "Rev. Mod. Phys.",
    volume = "35",
    pages = "335",
    year = "1963"
}

@article{Dombey:1969wk,
    author = "Dombey, Norman",
    title = "{Scattering of polarized leptons at high energy}",
    doi = "10.1103/RevModPhys.41.236",
    journal = "Rev. Mod. Phys.",
    volume = "41",
    pages = "236--246",
    year = "1969"
}

@article{Akhiezer:1973xbf,
    author = "Akhiezer, A. I. and Rekalo, Mikhail. P.",
    title = "{Polarization effects in the scattering of leptons by hadrons}",
    journal = "Fiz. Elem. Chast. Atom. Yadra",
    volume = "4",
    pages = "662--688",
    year = "1973"
}

@article{Drechsel:1989ab,
    author = "Drechsel, D. and Giannini, M. M.",
    title = "{ELECTRON SCATTERING OFF NUCLEI}",
    doi = "10.1088/0034-4885/52/9/002",
    journal = "Rept. Prog. Phys.",
    volume = "52",
    pages = "1083--1163",
    year = "1989"
}

@article{Boffi:1993gs,
    author = "Boffi, S. and Giusti, C. and Pacati, F. D.",
    title = "{Nuclear response in electromagnetic interactions with complex nuclei}",
    doi = "10.1016/0370-1573(93)90132-W",
    journal = "Phys. Rept.",
    volume = "226",
    pages = "1--101",
    year = "1993"
}

@article{Blomqvist:1998xn,
    author = "Blomqvist, K. I. and others",
    title = "{The three-spectrometer facility at the Mainz microtron MAMI}",
    doi = "10.1016/S0168-9002(97)01133-9",
    journal = "Nucl. Instrum. Meth. A",
    volume = "403",
    pages = "263--301",
    year = "1998"
}

@article{Leemann:2001dg,
    author = "Leemann, C. W. and Douglas, D. R. and Krafft, G. A.",
    title = "{The Continuous Electron Beam Accelerator Facility: CEBAF at the Jefferson Laboratory}",
    reportNumber = "JLAB-ACC-01-31",
    doi = "10.1146/annurev.nucl.51.101701.132327",
    journal = "Ann. Rev. Nucl. Part. Sci.",
    volume = "51",
    pages = "413--450",
    year = "2001"
}

@article{Arrington:2006zm,
    author = "Arrington, J. and Roberts, C. D. and Zanotti, J. M.",
    title = "{Nucleon electromagnetic form-factors}",
    eprint = "nucl-th/0611050",
    archivePrefix = "arXiv",
    reportNumber = "ANL-PHY-11657-TH-2006",
    doi = "10.1088/0954-3899/34/7/S03",
    journal = "J. Phys. G",
    volume = "34",
    pages = "S23--S52",
    year = "2007"
}

@article{Zhan:2011ji,
    author = "Zhan, X. and others",
    title = "{High-Precision Measurement of the Proton Elastic Form Factor Ratio $\mu_pG_E/G_M$ at low $Q^2$}",
    eprint = "1102.0318",
    archivePrefix = "arXiv",
    primaryClass = "nucl-ex",
    reportNumber = "JLAB-PHY-11-1311",
    doi = "10.1016/j.physletb.2011.10.002",
    journal = "Phys. Lett. B",
    volume = "705",
    pages = "59--64",
    year = "2011"
}

@article{Abrahamyan:2012gp,
    author = "Abrahamyan, S. and others",
    title = "{Measurement of the Neutron Radius of 208Pb Through Parity-Violation in Electron Scattering}",
    eprint = "1201.2568",
    archivePrefix = "arXiv",
    primaryClass = "nucl-ex",
    reportNumber = "JLAB-PHY-12-1480",
    doi = "10.1103/PhysRevLett.108.112502",
    journal = "Phys. Rev. Lett.",
    volume = "108",
    pages = "112502",
    year = "2012"
}

@article{Accardi:2012qut,
    author = "Accardi, A. and others",
    editor = "Deshpande, A. and Meziani, Z. E. and Qiu, J. W.",
    title = "{Electron Ion Collider: The Next QCD Frontier}: {Understanding the glue that binds us all}",
    eprint = "1212.1701",
    archivePrefix = "arXiv",
    primaryClass = "nucl-ex",
    reportNumber = "BNL-98815-2012-JA, JLAB-PHY-12-1652",
    doi = "10.1140/epja/i2016-16268-9",
    journal = "Eur. Phys. J. A",
    volume = "52",
    number = "9",
    pages = "268",
    year = "2016"
}

@article{Qweak:2013zxf,
    author = "Androic, D. and others",
    collaboration = "Qweak",
    title = "{First Determination of the Weak Charge of the Proton}",
    eprint = "1307.5275",
    archivePrefix = "arXiv",
    primaryClass = "nucl-ex",
    reportNumber = "JLAB-PHY-13-1756",
    doi = "10.1103/PhysRevLett.111.141803",
    journal = "Phys. Rev. Lett.",
    volume = "111",
    number = "14",
    pages = "141803",
    year = "2013"
}

@article{MUSE:2013uhu,
    author = "Gilman, R. and others",
    collaboration = "MUSE",
    title = "{Studying the Proton ''Radius'' Puzzle with {\textbackslash}mu p Elastic Scattering}",
    eprint = "1303.2160",
    archivePrefix = "arXiv",
    primaryClass = "nucl-ex",
    month = "3",
    year = "2013"
}

@article{Mosel:2016cwa,
    author = "Mosel, Ulrich",
    title = "{Neutrino Interactions with Nucleons and Nuclei: Importance for Long-Baseline Experiments}",
    eprint = "1602.00696",
    archivePrefix = "arXiv",
    primaryClass = "nucl-th",
    doi = "10.1146/annurev-nucl-102115-044720",
    journal = "Ann. Rev. Nucl. Part. Sci.",
    volume = "66",
    pages = "171--195",
    year = "2016"
}

@article{Aschenauer:2017jsk,
    author = "Aschenauer, E. C. and Fazio, S. and Lee, J. H. and Mantysaari, H. and Page, B. S. and Schenke, B. and Ullrich, T. and Venugopalan, R. and Zurita, P.",
    title = "{The electron{\textendash}ion collider: assessing the energy dependence of key measurements}",
    eprint = "1708.01527",
    archivePrefix = "arXiv",
    primaryClass = "nucl-ex",
    reportNumber = "BNL-114111-2017",
    doi = "10.1088/1361-6633/aaf216",
    journal = "Rept. Prog. Phys.",
    volume = "82",
    number = "2",
    pages = "024301",
    year = "2019"
}

@article{Adams:2018pwt,
    author = "Adams, B. and others",
    title = "{Letter of Intent: A New QCD facility at the M2 beam line of the CERN SPS (COMPASS++/AMBER)}",
    eprint = "1808.00848",
    archivePrefix = "arXiv",
    primaryClass = "hep-ex",
    reportNumber = "CERN-SPSC-2019-003, SPSC-I-250",
    month = "8",
    year = "2018"
}

@article{Xiong:2019umf,
    author = "Xiong, W. and others",
    title = "{A small proton charge radius from an electron{\textendash}proton scattering experiment}",
    doi = "10.1038/s41586-019-1721-2",
    journal = "Nature",
    volume = "575",
    number = "7781",
    pages = "147--150",
    year = "2019"
}

@article{Erler:2004in,
    author = "Erler, Jens and Ramsey-Musolf, Michael J.",
    title = "{The Weak mixing angle at low energies}",
    eprint = "hep-ph/0409169",
    archivePrefix = "arXiv",
    reportNumber = "FT-2004-03, CALTECH-MAP-300",
    doi = "10.1103/PhysRevD.72.073003",
    journal = "Phys. Rev. D",
    volume = "72",
    pages = "073003",
    year = "2005"
}

@article{SLACE158:2005uay,
    author = "Anthony, P. L. and others",
    collaboration = "SLAC E158",
    title = "{Precision measurement of the weak mixing angle in Moller scattering}",
    eprint = "hep-ex/0504049",
    archivePrefix = "arXiv",
    reportNumber = "SLAC-PUB-11149",
    doi = "10.1103/PhysRevLett.95.081601",
    journal = "Phys. Rev. Lett.",
    volume = "95",
    pages = "081601",
    year = "2005"
}

@article{Kumar:2013yoa,
    author = "Kumar, K. S. and Mantry, Sonny and Marciano, W. J. and Souder, P. A.",
    title = "{Low Energy Measurements of the Weak Mixing Angle}",
    eprint = "1302.6263",
    archivePrefix = "arXiv",
    primaryClass = "hep-ex",
    doi = "10.1146/annurev-nucl-102212-170556",
    journal = "Ann. Rev. Nucl. Part. Sci.",
    volume = "63",
    pages = "237--267",
    year = "2013"
}

@article{Becker:2018ggl,
    author = "Becker, Dominik and others",
    title = "{The P2 experiment}",
    eprint = "1802.04759",
    archivePrefix = "arXiv",
    primaryClass = "nucl-ex",
    doi = "10.1140/epja/i2018-12611-6",
    journal = "Eur. Phys. J. A",
    volume = "54",
    number = "11",
    pages = "208",
    year = "2018"
}

@article{MINOS:2011amj,
    author = "Adamson, P. and others",
    collaboration = "MINOS",
    title = "{Improved search for muon-neutrino to electron-neutrino oscillations in MINOS}",
    eprint = "1108.0015",
    archivePrefix = "arXiv",
    primaryClass = "hep-ex",
    reportNumber = "FERMILAB-PUB-11-351-PPD, BNL-96120-2011-JA",
    doi = "10.1103/PhysRevLett.107.181802",
    journal = "Phys. Rev. Lett.",
    volume = "107",
    pages = "181802",
    year = "2011"
}

@article{T2K:2011qtm,
    author = "Abe, K. and others",
    collaboration = "T2K",
    title = "{The T2K Experiment}",
    eprint = "1106.1238",
    archivePrefix = "arXiv",
    primaryClass = "physics.ins-det",
    doi = "10.1016/j.nima.2011.06.067",
    journal = "Nucl. Instrum. Meth. A",
    volume = "659",
    pages = "106--135",
    year = "2011"
}

@article{Hyper-KamiokandeProto-:2015xww,
    author = "Abe, K. and others",
    collaboration = "Hyper-Kamiokande Proto-",
    title = "{Physics potential of a long-baseline neutrino oscillation experiment using a J-PARC neutrino beam and Hyper-Kamiokande}",
    eprint = "1502.05199",
    archivePrefix = "arXiv",
    primaryClass = "hep-ex",
    doi = "10.1093/ptep/ptv061",
    journal = "PTEP",
    volume = "2015",
    pages = "053C02",
    year = "2015"
}

@article{T2K:2019bcf,
    author = "Abe, K. and others",
    collaboration = "T2K",
    title = "{Constraint on the matter{\textendash}antimatter symmetry-violating phase in neutrino oscillations}",
    eprint = "1910.03887",
    archivePrefix = "arXiv",
    primaryClass = "hep-ex",
    doi = "10.1038/s41586-020-2177-0",
    journal = "Nature",
    volume = "580",
    number = "7803",
    pages = "339--344",
    year = "2020",
    note = "[Erratum: Nature 583, E16 (2020)]"
}

@article{NOvA:2019cyt,
    author = "Acero, M. A. and others",
    collaboration = "NOvA",
    title = "{First Measurement of Neutrino Oscillation Parameters using Neutrinos and Antineutrinos by NOvA}",
    eprint = "1906.04907",
    archivePrefix = "arXiv",
    primaryClass = "hep-ex",
    reportNumber = "FERMILAB-PUB-19-272-ND",
    doi = "10.1103/PhysRevLett.123.151803",
    journal = "Phys. Rev. Lett.",
    volume = "123",
    number = "15",
    pages = "151803",
    year = "2019"
}

@article{DUNE:2020ypp,
    author = "Abi, Babak and others",
    collaboration = "DUNE",
    title = "{Deep Underground Neutrino Experiment (DUNE), Far Detector Technical Design Report, Volume II: DUNE Physics}",
    eprint = "2002.03005",
    archivePrefix = "arXiv",
    primaryClass = "hep-ex",
    reportNumber = "FERMILAB-PUB-20-025-ND, FERMILAB-DESIGN-2020-02",
    doi = "10.2172/1599307",
    month = "2",
    year = "2020"
}

@article{Perdrisat:2006hj,
    author = "Perdrisat, C. F. and Punjabi, V. and Vanderhaeghen, M.",
    title = "{Nucleon Electromagnetic Form Factors}",
    eprint = "hep-ph/0612014",
    archivePrefix = "arXiv",
    reportNumber = "WM-06-115, JLAB-THY-06-595",
    doi = "10.1016/j.ppnp.2007.05.001",
    journal = "Prog. Part. Nucl. Phys.",
    volume = "59",
    pages = "694--764",
    year = "2007"
}

@article{JeffersonLabHallA:2022cit,
    author = "Jiang, L. and others",
    collaboration = "Jefferson Lab Hall A",
    title = "{Determination of the argon spectral function from (e,e'p) data}",
    eprint = "2203.01748",
    archivePrefix = "arXiv",
    primaryClass = "nucl-ex",
    reportNumber = "JLAB-PHY-22-3575, SLAC-PUB-17650",
    doi = "10.1103/PhysRevD.105.112002",
    journal = "Phys. Rev. D",
    volume = "105",
    number = "11",
    pages = "112002",
    year = "2022"
}

@article{JeffersonLabHallA:2022ljj,
    author = "Jiang, L. and others",
    collaboration = "Jefferson Lab Hall A",
    title = "{Determination of the titanium spectral function from (e,{\,}e'p) data}",
    eprint = "2209.14108",
    archivePrefix = "arXiv",
    primaryClass = "nucl-ex",
    doi = "10.1103/PhysRevD.107.012005",
    journal = "Phys. Rev. D",
    volume = "107",
    number = "1",
    pages = "012005",
    year = "2023"
}

@article{DeRujula:1979grv,
    author = "De Rujula, A. and Petronzio, R. and Savoy-Navarro, A.",
    title = "{Radiative Corrections to High-Energy Neutrino Scattering}",
    reportNumber = "CERN-TH-2593",
    doi = "10.1016/0550-3213(79)90039-7",
    journal = "Nucl. Phys. B",
    volume = "154",
    pages = "394--426",
    year = "1979"
}

@article{Day:2012gb,
    author = "Day, Melanie and McFarland, Kevin S.",
    title = "{Differences in Quasi-Elastic Cross-Sections of Muon and Electron Neutrinos}",
    eprint = "1206.6745",
    archivePrefix = "arXiv",
    primaryClass = "hep-ph",
    reportNumber = "FERMILAB-PUB-12-314-PPD",
    doi = "10.1103/PhysRevD.86.053003",
    journal = "Phys. Rev. D",
    volume = "86",
    pages = "053003",
    year = "2012"
}

@article{Tomalak:2021hec,
    author = "Tomalak, Oleksandr and Chen, Qing and Hill, Richard J. and McFarland, Kevin S.",
    title = "{QED radiative corrections for accelerator neutrinos}",
    eprint = "2105.07939",
    archivePrefix = "arXiv",
    primaryClass = "hep-ph",
    reportNumber = "FERMILAB-PUB-22-677-V, FERMILAB-PUB-21-232-T, LA-UR-21-27844, USTC-ICTS/PCFT-21-32",
    doi = "10.1038/s41467-022-32974-x",
    journal = "Nature Commun.",
    volume = "13",
    number = "1",
    pages = "5286",
    year = "2022"
}

@article{Tomalak:2022xup,
    author = "Tomalak, Oleksandr and Chen, Qing and Hill, Richard J. and McFarland, Kevin S. and Wret, Clarence",
    title = "{Theory of QED radiative corrections to neutrino scattering at accelerator energies}",
    eprint = "2204.11379",
    archivePrefix = "arXiv",
    primaryClass = "hep-ph",
    reportNumber = "FERMILAB-PUB-21-378-T, LA-UR-21-30632, USTC-ICTS/PCFT-22-13",
    doi = "10.1103/PhysRevD.106.093006",
    journal = "Phys. Rev. D",
    volume = "106",
    number = "9",
    pages = "093006",
    year = "2022"
}

@article{Calva-Tellez:1978ufm,
    author = "Calva-Tellez, E. and Yennie, D. R.",
    title = "{Coulomb Corrections to Deep Inelastic Electron or Muon Scattering From Nuclei}",
    reportNumber = "CLNS-407",
    doi = "10.1103/PhysRevD.20.105",
    journal = "Phys. Rev. D",
    volume = "20",
    pages = "105",
    year = "1979"
}

@article{Hill:2023bfh,
    author = "Hill, Richard J. and Plestid, Ryan",
    title = "{All orders factorization and the Coulomb problem}",
    eprint = "2309.15929",
    archivePrefix = "arXiv",
    primaryClass = "hep-ph",
    reportNumber = "CALT-TH-2023-034, FERMILAB-PUB-23-454-T",
    doi = "10.1103/PhysRevD.109.056006",
    journal = "Phys. Rev. D",
    volume = "109",
    number = "5",
    pages = "056006",
    year = "2024"
}

@article{Crowe:2026lky,
    author = "Crowe, Daniel and Hasan, Syed Mehedi and Wackeroth, Doreen",
    title = "{Radiative Corrections to Elastic Lepton-Proton Scattering with Focus on Two-Photon-Exchange Diagrams}",
    eprint = "2605.31123",
    archivePrefix = "arXiv",
    primaryClass = "hep-ph",
    month = "5",
    year = "2026"
}

@article{Kasahara:1985ke,
    author = "Kasahara, K.",
    title = "{EXPERIMENTAL EXAMINATION OF THE LANDAU-POMERANCHUK-MIGDAL EFFECT BY HIGH-ENERGY ELECTROMAGNETIC CASCADE SHOWERS IN LEAD}",
    doi = "10.1103/PhysRevD.31.2737",
    journal = "Phys. Rev. D",
    volume = "31",
    pages = "2737--2747",
    year = "1985"
}

@article{Anthony:1995fs,
    author = "Anthony, P. L. and others",
    title = "{An Accurate measurement of the Landau-Pomeranchuk-Migdal effect}",
    reportNumber = "SLAC-PUB-6796, SLAC-PUB-95-6796, LBL-37178",
    doi = "10.1103/PhysRevLett.75.1949",
    journal = "Phys. Rev. Lett.",
    volume = "75",
    pages = "1949--1952",
    year = "1995"
}

@article{SLAC-E-146:1997hnd,
    author = "Anthony, P. L. and others",
    collaboration = "SLAC-E-146",
    title = "{Bremsstrahlung suppression due to the LPM and dielectric effects in a variety of materials}",
    eprint = "hep-ex/9703016",
    archivePrefix = "arXiv",
    reportNumber = "SLAC-PUB-7413, LBL-40054, LBNL-40054, SLAC--PUB--7413, LBNL--40054",
    doi = "10.1103/PhysRevD.56.1373",
    journal = "Phys. Rev. D",
    volume = "56",
    pages = "1373--1390",
    year = "1997"
}

@article{Hansen:2004ti,
    author = "Hansen, H. D. and Uggerhj, Ulrik Ingerslev and Biino, C. and Ballestrero, S. and Mangiarotti, A. and Sona, P. and Ketel, T. J. and Vilakazi, Z. Z.",
    title = "{Landau-Pomeranchuk-Migdal effect for multihundred GeV electrons}",
    doi = "10.1103/PhysRevD.69.032001",
    journal = "Phys. Rev. D",
    volume = "69",
    pages = "032001",
    year = "2004"
}

@article{CERNNA63:2013ahd,
    author = "Andersen, K. K. and Andersen, S. L. and Esberg, J. and Knudsen, H. and Mikkelsen, R. E. and Uggerh{\o}j, U. I. and Wistisen, T. N. and Sona, P. and Mangiarotti, A. and Ketel, T. J.",
    collaboration = "CERN NA63",
    title = "{Experimental investigation of the Landau-Pomeranchuk-Migdal effect in low-Z targets}",
    eprint = "1309.5765",
    archivePrefix = "arXiv",
    primaryClass = "hep-ex",
    doi = "10.1103/PhysRevD.88.072007",
    journal = "Phys. Rev. D",
    volume = "88",
    number = "7",
    pages = "072007",
    year = "2013"
}

@misc{nds_charge_radii,
  url = {https://www-nds.iaea.org/radii/},
  author = {Marinova, Krassimira and Angeli, Istvan},
  title = {Nuclear Charge Radii},
  publisher = {Nuclear Data Services},
  year = {2022}
}

@article{Bhattacharya:2025pje,
    author = "Bhattacharya, Shohini and Tomalak, Oleksandr and Vitev, Ivan",
    title = "{QED nuclear medium effects at EIC energies}",
    eprint = "2502.06943",
    archivePrefix = "arXiv",
    primaryClass = "nucl-th",
    reportNumber = "LA-UR-24-25198, FERMILAB-PUB-24-0529-T",
    doi = "10.1103/PhysRevD.112.033001",
    journal = "Phys. Rev. D",
    volume = "112",
    number = "3",
    pages = "033001",
    year = "2025"
}

@article{Landau:1953um,
    author = "Landau, L. D. and Pomeranchuk, I.",
    title = "{Limits of applicability of the theory of bremsstrahlung electrons and pair production at high-energies}",
    journal = "Dokl. Akad. Nauk Ser. Fiz.",
    volume = "92",
    pages = "535--536",
    year = "1953"
}

@article{Landau:1953gr,
    author = "Landau, L. D. and Pomeranchuk, I.",
    title = "{Electron cascade process at very high-energies}",
    journal = "Dokl. Akad. Nauk Ser. Fiz.",
    volume = "92",
    pages = "735--738",
    year = "1953"
}

@article{Migdal:1956tc,
    author = "Migdal, A. B.",
    title = "{Bremsstrahlung and Pair Production at High Energies in Condensed Media}",
    doi = "10.1103/PhysRev.103.1811",
    journal = "Phys. Rev.",
    volume = "103",
    pages = "1811--1820",
    year = "1956"
}

@article{Baier:1996vi,
    author = "Baier, R. and Dokshitzer, Yuri L. and Mueller, Alfred H. and Peigne, S. and Schiff, D.",
    title = "{The Landau-Pomeranchuk-Migdal effect in QED}",
    eprint = "hep-ph/9604327",
    archivePrefix = "arXiv",
    reportNumber = "BI-TP-95-40, CERN-TH-96-14, CUTP-724, LPTHE-ORSAY-95-84",
    doi = "10.1016/0550-3213(96)00426-9",
    journal = "Nucl. Phys. B",
    volume = "478",
    pages = "577--597",
    year = "1996"
}

@article{Zakharov:1996fv,
    author = "Zakharov, B. G.",
    title = "{Fully quantum treatment of the Landau-Pomeranchuk-Migdal effect in QED and QCD}",
    eprint = "hep-ph/9607440",
    archivePrefix = "arXiv",
    doi = "10.1134/1.567126",
    journal = "JETP Lett.",
    volume = "63",
    pages = "952--957",
    year = "1996"
}

@article{Arnold:2018fjr,
    author = "Arnold, Peter and Iqbal, Shahin and Rase, Tanner",
    title = "{Strong- vs. weak-coupling pictures of jet quenching: a dry run using QED}",
    eprint = "1810.06578",
    archivePrefix = "arXiv",
    primaryClass = "hep-ph",
    doi = "10.1007/JHEP05(2019)004",
    journal = "JHEP",
    volume = "05",
    pages = "004",
    year = "2019"
}

@article{Arnold:2025dqj,
    author = "Arnold, Peter and Bautista, Joshua and Elgedawy, Omar and Iqbal, Shahin",
    title = "{Revisiting extremely high energy QED bremsstrahlung in matter: large modifications to the LPM effect}",
    eprint = "2508.21120",
    archivePrefix = "arXiv",
    primaryClass = "hep-ph",
    doi = "10.1007/JHEP03(2026)015",
    journal = "JHEP",
    volume = "03",
    pages = "015",
    year = "2026"
}

@article{Arnold:2026lwd,
    author = "Arnold, Peter and Bautista, Joshua and Elgedawy, Omar and Iqbal, Shahin",
    title = "{Extremely high-energy bremsstrahlung in matter}",
    eprint = "2604.18685",
    archivePrefix = "arXiv",
    primaryClass = "hep-ph",
    reportNumber = "CPHT-RR013.042026, CERN-TH-2026-091",
    month = "4",
    year = "2026"
}

@article{Arnold:2026wjy,
    author = "Arnold, Peter and Bautista, Joshua and Elgedawy, Omar and Iqbal, Shahin",
    title = "{Calculating extremely high energy bremsstrahlung in matter}",
    eprint = "2605.03002",
    archivePrefix = "arXiv",
    primaryClass = "hep-ph",
    reportNumber = "CPHT-RR014.042026 and CERN-TH-2026-090",
    month = "5",
    year = "2026"
}

@article{Bertulani:1987tz,
    author = "Bertulani, Carlos A. and Baur, Gerhard",
    title = "{Electromagnetic Processes in Relativistic Heavy Ion Collisions}",
    reportNumber = "JUL-2163",
    doi = "10.1016/0370-1573(88)90142-1",
    journal = "Phys. Rept.",
    volume = "163",
    pages = "299",
    year = "1988"
}

@article{Vidovic:1992ik,
    author = "Vidovic, M. and Greiner, M. and Best, C. and Soff, G.",
    title = "{Impact parameter dependence of the electromagnetic particle production in ultrarelativistic heavy ion collisions}",
    reportNumber = "GSI-92-52",
    doi = "10.1103/PhysRevC.47.2308",
    journal = "Phys. Rev. C",
    volume = "47",
    pages = "2308--2319",
    year = "1993"
}

@article{Olsen:2003mj,
    author = "Olsen, Haakon A.",
    title = "{Differential bremsstrahlung and pair production cross-sections at high-energies}",
    doi = "10.1103/PhysRevD.68.033008",
    journal = "Phys. Rev. D",
    volume = "68",
    pages = "033008",
    year = "2003"
}

@article{Lee:2004ina,
    author = "Lee, R. N. and Milstein, A. I. and Strakhovenko, V. M. and Schwarz, O. Ya.",
    title = "{Coulomb corrections to bremsstrahlung in electric field of heavy atom at high energies}",
    eprint = "hep-ph/0404224",
    archivePrefix = "arXiv",
    doi = "10.1134/1.1866193",
    journal = "J. Exp. Theor. Phys.",
    volume = "100",
    number = "1",
    pages = "1--13",
    year = "2005"
}

@article{Sandrock:2018ivj,
    author = "Sandrock, A. and Rhode, W.",
    title = "{Coulomb corrections to the bremsstrahlung and electron pair production cross section of high-energy muons on extended nuclei}",
    eprint = "1807.08475",
    archivePrefix = "arXiv",
    primaryClass = "hep-ph",
    month = "7",
    year = "2018"
}

@article{LlewellynSmith:1971uhs,
    author = "Llewellyn Smith, C. H.",
    title = "{Neutrino Reactions at Accelerator Energies}",
    reportNumber = "SLAC-PUB-0958",
    doi = "10.1016/0370-1573(72)90010-5",
    journal = "Phys. Rept.",
    volume = "3",
    pages = "261--379",
    year = "1972"
}

@article{Engel:1997fy,
    author = "Engel, Jonathan",
    title = "{Approximate treatment of lepton distortion in charged current neutrino scattering from nuclei}",
    eprint = "nucl-th/9711045",
    archivePrefix = "arXiv",
    doi = "10.1103/PhysRevC.57.2004",
    journal = "Phys. Rev. C",
    volume = "57",
    pages = "2004--2009",
    year = "1998"
}

@article{Tjon:2006qe,
    author = "Tjon, J. A. and Wallace, S. J.",
    title = "{Coulomb corrections in quasi-elastic scattering based on the eikonal expansion for electron wave functions}",
    eprint = "nucl-th/0610115",
    archivePrefix = "arXiv",
    reportNumber = "DOE-ER-40762-373",
    doi = "10.1103/PhysRevC.74.064602",
    journal = "Phys. Rev. C",
    volume = "74",
    pages = "064602",
    year = "2006"
}

@article{Tomalak:2024lme,
    author = "Tomalak, Oleksandr and Vitev, Ivan",
    title = "{Medium-induced photon bremsstrahlung in neutrino-nucleus, antineutrino-nucleus, and electron-nucleus scattering from multiple QED interactions}",
    eprint = "2402.16851",
    archivePrefix = "arXiv",
    primaryClass = "hep-ph",
    reportNumber = "LA-UR-23-30368",
    doi = "10.1103/PhysRevD.109.073010",
    journal = "Phys. Rev. D",
    volume = "109",
    number = "7",
    pages = "073010",
    year = "2024"
}

@article{Yennie:1961ad,
    author = "Yennie, D. R. and Frautschi, Steven C. and Suura, H.",
    title = "{The infrared divergence phenomena and high-energy processes}",
    doi = "10.1016/0003-4916(61)90151-8",
    journal = "Annals Phys.",
    volume = "13",
    pages = "379--452",
    year = "1961"
}

@article{Mo:1968cg,
    author = "Mo, Luke W. and Tsai, Yung-Su",
    title = "{Radiative Corrections to Elastic and Inelastic e p and mu p Scattering}",
    reportNumber = "SLAC-PUB-0380",
    doi = "10.1103/RevModPhys.41.205",
    journal = "Rev. Mod. Phys.",
    volume = "41",
    pages = "205--235",
    year = "1969"
}

@article{Maximon:2000hm,
    author = "Maximon, L. C. and Tjon, J. A.",
    title = "{Radiative corrections to electron proton scattering}",
    eprint = "nucl-th/0002058",
    archivePrefix = "arXiv",
    doi = "10.1103/PhysRevC.62.054320",
    journal = "Phys. Rev. C",
    volume = "62",
    pages = "054320",
    year = "2000"
}

@article{Vanderhaeghen:2000ws,
    author = "Vanderhaeghen, M. and Friedrich, J. M. and Lhuillier, D. and Marchand, D. and Van Hoorebeke, L. and Van de Wiele, J.",
    title = "{QED radiative corrections to virtual Compton scattering}",
    eprint = "hep-ph/0001100",
    archivePrefix = "arXiv",
    doi = "10.1103/PhysRevC.62.025501",
    journal = "Phys. Rev. C",
    volume = "62",
    pages = "025501",
    year = "2000"
}

@article{Gramolin:2014pva,
    author = "Gramolin, A. V. and Fadin, V. S. and Feldman, A. L. and Gerasimov, R. E. and Nikolenko, D. M. and Rachek, I. A. and Toporkov, D. K.",
    title = "{A new event generator for the elastic scattering of charged leptons on protons}",
    eprint = "1401.2959",
    archivePrefix = "arXiv",
    primaryClass = "nucl-ex",
    doi = "10.1088/0954-3899/41/11/115001",
    journal = "J. Phys. G",
    volume = "41",
    number = "11",
    pages = "115001",
    year = "2014"
}

@article{Liu:2020rvc,
    author = "Liu, Tianbo and Melnitchouk, W. and Qiu, Jian-Wei and Sato, N.",
    title = "{Factorized approach to radiative corrections for inelastic lepton-hadron collisions}",
    eprint = "2008.02895",
    archivePrefix = "arXiv",
    primaryClass = "hep-ph",
    reportNumber = "JLAB-THY-20-3233",
    doi = "10.1103/PhysRevD.104.094033",
    journal = "Phys. Rev. D",
    volume = "104",
    number = "9",
    pages = "094033",
    year = "2021"
}

@article{Afanasev:2023gev,
    author = "Afanasev, Andrei and others",
    title = "{Radiative corrections: from medium to high energy experiments}",
    eprint = "2306.14578",
    archivePrefix = "arXiv",
    primaryClass = "hep-ph",
    doi = "10.1140/epja/s10050-024-01281-y",
    journal = "Eur. Phys. J. A",
    volume = "60",
    number = "4",
    pages = "91",
    year = "2024"
}

@article{Tuchin:2013eya,
    author = "Tuchin, Kirill",
    title = "{Coulomb corrections to photon and dilepton production in high energy $pA$ collisions}",
    eprint = "1311.1124",
    archivePrefix = "arXiv",
    primaryClass = "hep-ph",
    doi = "10.1103/PhysRevC.89.024904",
    journal = "Phys. Rev. C",
    volume = "89",
    number = "2",
    pages = "024904",
    year = "2014"
}

@article{Sun:2020ygb,
    author = "Sun, Ze-hao and Zheng, Du-xin and Zhou, Jian and Zhou, Ya-jin",
    title = "{Studying Coulomb correction at EIC and EicC}",
    eprint = "2002.07373",
    archivePrefix = "arXiv",
    primaryClass = "hep-ph",
    doi = "10.1016/j.physletb.2020.135679",
    journal = "Phys. Lett. B",
    volume = "808",
    pages = "135679",
    year = "2020"
}

@article{Idilbi:2008vm,
    author = "Idilbi, Ahmad and Majumder, Abhijit",
    title = "{Extending Soft-Collinear-Effective-Theory to describe hard jets in dense QCD media}",
    eprint = "0808.1087",
    archivePrefix = "arXiv",
    primaryClass = "hep-ph",
    doi = "10.1103/PhysRevD.80.054022",
    journal = "Phys. Rev. D",
    volume = "80",
    pages = "054022",
    year = "2009"
}

@article{Rothstein:2016bsq,
    author = "Rothstein, Ira Z. and Stewart, Iain W.",
    title = "{An Effective Field Theory for Forward Scattering and Factorization Violation}",
    eprint = "1601.04695",
    archivePrefix = "arXiv",
    primaryClass = "hep-ph",
    reportNumber = "MIT-CTP-4655, MIT-CTP 4655",
    doi = "10.1007/JHEP08(2016)025",
    journal = "JHEP",
    volume = "08",
    pages = "025",
    year = "2016"
}

@article{Hill:2019xqk,
    author = "Hill, Richard J. and Tomalak, Oleksandr",
    title = "{On the effective theory of neutrino-electron and neutrino-quark interactions}",
    eprint = "1911.01493",
    archivePrefix = "arXiv",
    primaryClass = "hep-ph",
    reportNumber = "FERMILAB-PUB-19-559-T",
    doi = "10.1016/j.physletb.2020.135466",
    journal = "Phys. Lett. B",
    volume = "805",
    pages = "135466",
    year = "2020"
}

@article{ParticleDataGroup:2024cfk,
    author = "Navas, S. and others",
    collaboration = "Particle Data Group",
    title = "{Review of particle physics}",
    doi = "10.1103/PhysRevD.110.030001",
    journal = "Phys. Rev. D",
    volume = "110",
    number = "3",
    pages = "030001",
    year = "2024"
}

@article{Kusina:2020lyz,
    author = "Kusina, A. and others",
    title = "{Impact of LHC vector boson production in heavy ion collisions on strange PDFs}",
    eprint = "2007.09100",
    archivePrefix = "arXiv",
    primaryClass = "hep-ph",
    reportNumber = "IFJPAN-IV-2020-4, KA-TP-07-2020, MS-TP-20-27, P3H-20-035,
  SMU-HEP-20-04",
    doi = "10.1140/epjc/s10052-020-08532-4",
    journal = "Eur. Phys. J. C",
    volume = "80",
    number = "10",
    pages = "968",
    year = "2020"
}

@article{Kovarik:2015cma,
    author = "Kovarik, K. and others",
    title = "{nCTEQ15 - Global analysis of nuclear parton distributions with uncertainties in the CTEQ framework}",
    eprint = "1509.00792",
    archivePrefix = "arXiv",
    primaryClass = "hep-ph",
    reportNumber = "LPSC-15-153, MS-TP-15-11, FERMILAB-PUB-15-375-ND-PPD-T",
    doi = "10.1103/PhysRevD.93.085037",
    journal = "Phys. Rev. D",
    volume = "93",
    number = "8",
    pages = "085037",
    year = "2016"
}

@book{Halzen:1984mc,
    author = "Halzen, F. and Martin, Alan D.",
    title = "{QUARKS AND LEPTONS: AN INTRODUCTORY COURSE IN MODERN PARTICLE PHYSICS}",
    isbn = "978-0-471-88741-6",
    publisher = "John Wiley and Sons",
    year = "1984"
}

@article{CTEQ:1993hwr,
    author = "Brock, Raymond and others",
    collaboration = "CTEQ",
    title = "{Handbook of perturbative QCD: Version 1.0}",
    reportNumber = "FERMILAB-PUB-93-094, ANL-HEP-PR-95-29",
    doi = "10.1103/RevModPhys.67.157",
    journal = "Rev. Mod. Phys.",
    volume = "67",
    pages = "157--248",
    year = "1995"
}

@article{Clark:2016jgm,
    author = "Clark, D. B. and Godat, E. and Olness, F. I.",
    title = "{ManeParse : A Mathematica  reader for Parton Distribution Functions}",
    eprint = "1605.08012",
    archivePrefix = "arXiv",
    primaryClass = "hep-ph",
    reportNumber = "NSF-KITP-16-032, SMU-HEP-16-05",
    doi = "10.1016/j.cpc.2017.03.004",
    journal = "Comput. Phys. Commun.",
    volume = "216",
    pages = "126--137",
    year = "2017"
}

@article{Ovanesyan:2011xy,
    author = "Ovanesyan, Grigory and Vitev, Ivan",
    title = "{An effective theory for jet propagation in dense QCD matter: jet broadening and medium-induced bremsstrahlung}",
    eprint = "1103.1074",
    archivePrefix = "arXiv",
    primaryClass = "hep-ph",
    doi = "10.1007/JHEP06(2011)080",
    journal = "JHEP",
    volume = "06",
    pages = "080",
    year = "2011"
}

@book{Jackson:1998nia,
    author = "Jackson, John David",
    title = "{Classical Electrodynamics}",
    isbn = "978-0-471-30932-1",
    publisher = "Wiley",
    year = "1998"
}

@article{JSSv047s02,
 title={DataGraph 3.0},
 volume={47},
 url={https://www.jstatsoft.org/index.php/jss/article/view/v047s02},
 doi={10.18637/jss.v047.s02},
 number={2},
 journal={Journal of Statistical Software},
 author={MacAskill, Michael R.},
 year={2012},
 pages={1–9}
}

@misc{Mathematica,
  author = {{Wolfram~Research{,}~Inc.}},
  title = {Mathematica, {V}ersion 12.2.0.0},
  note = {{Champaign, IL}},
  year = {2022}
}

@article{Mertig:1990an,
    author = "Mertig, R. and Bohm, M. and Denner, Ansgar",
    title = "{FEYN CALC: Computer algebraic calculation of Feynman amplitudes}",
    reportNumber = "PRINT-90-0639 (WURZBURG)",
    doi = "10.1016/0010-4655(91)90130-D",
    journal = "Comput. Phys. Commun.",
    volume = "64",
    pages = "345--359",
    year = "1991"
}

@article{Shtabovenko:2016sxi,
    author = "Shtabovenko, Vladyslav and Mertig, Rolf and Orellana, Frederik",
    title = "{New Developments in FeynCalc 9.0}",
    eprint = "1601.01167",
    archivePrefix = "arXiv",
    primaryClass = "hep-ph",
    reportNumber = "TUM-EFT-71-15",
    doi = "10.1016/j.cpc.2016.06.008",
    journal = "Comput. Phys. Commun.",
    volume = "207",
    pages = "432--444",
    year = "2016"
}

@article{A1:2013fsc,
    author = "Bernauer, J. C. and others",
    collaboration = "A1",
    title = "{Electric and magnetic form factors of the proton}",
    eprint = "1307.6227",
    archivePrefix = "arXiv",
    primaryClass = "nucl-ex",
    doi = "10.1103/PhysRevC.90.015206",
    journal = "Phys. Rev. C",
    volume = "90",
    number = "1",
    pages = "015206",
    year = "2014"
}

@article{A1:2010nsl,
    author = "Bernauer, J. C. and others",
    collaboration = "A1",
    title = "{High-precision determination of the electric and magnetic form factors of the proton}",
    eprint = "1007.5076",
    archivePrefix = "arXiv",
    primaryClass = "nucl-ex",
    doi = "10.1103/PhysRevLett.105.242001",
    journal = "Phys. Rev. Lett.",
    volume = "105",
    pages = "242001",
    year = "2010"
}

@article{Meyer:2016oeg,
    author = "Meyer, Aaron S. and Betancourt, Minerba and Gran, Richard and Hill, Richard J.",
    title = "{Deuterium target data for precision neutrino-nucleus cross sections}",
    eprint = "1603.03048",
    archivePrefix = "arXiv",
    primaryClass = "hep-ph",
    reportNumber = "FERMILAB-PUB-16-185-ND-T",
    doi = "10.1103/PhysRevD.93.113015",
    journal = "Phys. Rev. D",
    volume = "93",
    number = "11",
    pages = "113015",
    year = "2016"
}

@article{Borah:2020gte,
    author = "Borah, Kaushik and Hill, Richard J. and Lee, Gabriel and Tomalak, Oleksandr",
    title = "{Parametrization and applications of the low-$Q^2$ nucleon vector form factors}",
    eprint = "2003.13640",
    archivePrefix = "arXiv",
    primaryClass = "hep-ph",
    reportNumber = "FERMILAB-PUB-20-124-T",
    doi = "10.1103/PhysRevD.102.074012",
    journal = "Phys. Rev. D",
    volume = "102",
    number = "7",
    pages = "074012",
    year = "2020"
}

@article{Willeke:2021ymc,
    author = "Willeke, Ferdinand",
    title = "{Electron Ion Collider Conceptual Design Report 2021}",
    reportNumber = "BNL-221006-2021-FORE",
    journal = "Conceptual Design Report",
    doi = "10.2172/1765663",
    month = "2",
    year = "2021"
}

@article{Anderle:2021wcy,
    author = "Anderle, Daniele P. and others",
    title = "{Electron-ion collider in China}",
    eprint = "2102.09222",
    archivePrefix = "arXiv",
    primaryClass = "nucl-ex",
    reportNumber = "Frontiers of Physics, Volume 16 Issue (6):64701, 2021",
    doi = "10.1007/s11467-021-1062-0",
    journal = "Front. Phys. (Beijing)",
    volume = "16",
    number = "6",
    pages = "64701",
    year = "2021"
}

@article{Tomalak:2022kjd,
    author = "Tomalak, Oleksandr and Vitev, Ivan",
    title = "{QED medium effects in (anti)neutrino-nucleus and electron-nucleus scattering: Elastic scattering on nucleons}",
    eprint = "2206.10637",
    archivePrefix = "arXiv",
    primaryClass = "nucl-th",
    reportNumber = "LA-UR-22-22741",
    doi = "10.1016/j.physletb.2022.137492",
    journal = "Phys. Lett. B",
    volume = "835",
    pages = "137492",
    year = "2022"
}

@article{Tomalak:2023kwl,
    author = "Tomalak, Oleksandr and Vitev, Ivan",
    title = "{Broadening of particle distributions in electron-, neutrino-, and antineutrino-nucleus scattering from QED interactions}",
    eprint = "2310.01414",
    archivePrefix = "arXiv",
    primaryClass = "hep-ph",
    reportNumber = "LA-UR-21-30814",
    doi = "10.1103/PhysRevD.108.093003",
    journal = "Phys. Rev. D",
    volume = "108",
    number = "9",
    pages = "093003",
    year = "2023"
}

@article{AbdulKhalek:2021gbh,
    author = "Abdul Khalek, R. and others",
    title = "{Science Requirements and Detector Concepts for the Electron-Ion Collider}: {EIC Yellow Report}",
    eprint = "2103.05419",
    archivePrefix = "arXiv",
    primaryClass = "physics.ins-det",
    reportNumber = "BNL-220990-2021-FORE, JLAB-PHY-21-3198, LA-UR-21-20953",
    doi = "10.1016/j.nuclphysa.2022.122447",
    journal = "Nucl. Phys. A",
    volume = "1026",
    pages = "122447",
    year = "2022"
}

@article{Gyulassy:2002yv,
    author = "Gyulassy, M. and Levai, P. and Vitev, I.",
    title = "{Reaction operator approach to multiple elastic scatterings}",
    eprint = "nucl-th/0201078",
    archivePrefix = "arXiv",
    doi = "10.1103/PhysRevD.66.014005",
    journal = "Phys. Rev. D",
    volume = "66",
    pages = "014005",
    year = "2002"
}

@article{Barata:2020rdn,
    author = "Barata, Jo\~ao and Mehtar-Tani, Yacine and Soto-Ontoso, Alba and Tywoniuk, Konrad",
    title = "{Revisiting transverse momentum broadening in dense QCD media}",
    eprint = "2009.13667",
    archivePrefix = "arXiv",
    primaryClass = "hep-ph",
    doi = "10.1103/PhysRevD.104.054047",
    journal = "Phys. Rev. D",
    volume = "104",
    number = "5",
    pages = "054047",
    year = "2021"
}

@article{Gyulassy:2000fs,
    author = "Gyulassy, M. and Levai, P. and Vitev, I.",
    title = "{NonAbelian energy loss at finite opacity}",
    eprint = "nucl-th/0005032",
    archivePrefix = "arXiv",
    reportNumber = "CU-TP-976",
    doi = "10.1103/PhysRevLett.85.5535",
    journal = "Phys. Rev. Lett.",
    volume = "85",
    pages = "5535--5538",
    year = "2000"
}

@article{Gyulassy:2000er,
    author = "Gyulassy, M. and Levai, P. and Vitev, I.",
    title = "{Reaction operator approach to nonAbelian energy loss}",
    eprint = "nucl-th/0006010",
    archivePrefix = "arXiv",
    reportNumber = "CU-TP-979",
    doi = "10.1016/S0550-3213(00)00652-0",
    journal = "Nucl. Phys. B",
    volume = "594",
    pages = "371--419",
    year = "2001"
}

@article{Wiedemann:2000za,
    author = "Wiedemann, Urs Achim",
    title = "{Gluon radiation off hard quarks in a nuclear environment: Opacity expansion}",
    eprint = "hep-ph/0005129",
    archivePrefix = "arXiv",
    doi = "10.1016/S0550-3213(00)00457-0",
    journal = "Nucl. Phys. B",
    volume = "588",
    pages = "303--344",
    year = "2000"
}

@article{Accardi:2009qv,
    author = "Accardi, A. and Arleo, F. and Brooks, W. K. and D'Enterria, David and Muccifora, V.",
    title = "{Parton Propagation and Fragmentation in QCD Matter}",
    eprint = "0907.3534",
    archivePrefix = "arXiv",
    primaryClass = "nucl-th",
    reportNumber = "JLAB-THY-09-1038",
    doi = "10.1393/ncr/i2009-10048-0",
    journal = "Riv. Nuovo Cim.",
    volume = "32",
    number = "9-10",
    pages = "439--554",
    year = "2009"
}

\end{document}